\documentclass{pasa}%

\usepackage{graphicx}

\newcommand{\gimic}        {\textsc{GIMIC\,}}
\newcommand{\gadget}      {\textsc{Gadget-3}}
\newcommand{\subfind}    {\textsc{SUBFIND\,}}
\newcommand{\sz}             {\hbox{$0\sigma$}\,}
\newcommand{\smm}        {\hbox{$-2\sigma$}\,}

\title[The diversity of assembly histories leading to disc galaxy formation in a $\Lambda$CDM model]{The diversity of assembly histories leading to disc galaxy formation in a $\Lambda$CDM model}

\author[Font et al.]{Andreea S. Font$^1$\thanks{email: A.S.Font@ljmu.ac.uk}, Ian G. McCarthy$^1$, Amandine M. C. Le Brun$^{2,3}$, Robert A. Crain$^{1,}$\ and Lee S. Kelvin$^1$ 
\affil{$^1$Astrophysics Research Institute, Liverpool John Moores University, 146 Brownlow Hill, Liverpool, L3 5RF, UK}%
\affil{$^2$IRFU, CEA, Universit{\'e} Paris-Saclay, F-91191 Gif-sur-Yvette, France}
\affil{$^3$Universit{\'e} Paris Diderot, AIM, Sorbonne Paris Cit{\'e}, CEA, CNRS, F-91191 Gif-sur-Yvette, France}
}%

\jid{PASA}
\doi{10.1017/pas.\the\year.xxx}
\jyear{\the\year}

\usepackage{aas_macros}
\usepackage{hyperref} 
\hypersetup{colorlinks,citecolor=blue,linkcolor=blue,urlcolor=blue}

\begin{document}

\begin{frontmatter}
\maketitle

\begin{abstract}
Typical disc galaxies forming in a $\Lambda$CDM cosmology encounter a violent environment, where they often experience mergers with satellites that can disrupt the discs.  However, disc galaxies are ubiquitous in the local Universe, suggesting that a quiescent formation is not necessary for their formation. Modern cosmological simulations in the context of the $\Lambda$CDM model can now obtain relatively realistic populations of disc galaxies.  In spite of this important success, it still remains to be clarified how discs manage to survive massive mergers and what are the various formation scenarios for disc galaxies.  Here we use a suite of high-resolution hydrodynamical simulations set in a $\Lambda$CDM cosmology to elucidate the fate of discs encountering mergers with systems several times more massive than their total stellar component, ${\rm M}_{\rm star, host}$ (comprising the majority of simulated disc galaxies since z=2).  For this, we extract a sample of approximately 100 disc galaxies and follow the changes in their post-merger morphologies, as tracked by their disc-to-total ratios (D/T). We also examine the relations between their present-day (kinematic) morphology, assembly history and gas fractions.  We find that approximately half of present-day disc galaxies underwent at least one merger with a satellite of total (dark matter, stars and gas) mass exceeding the host system's stellar mass, a third had mergers with satellites of mass exceeding 3 times the host's stellar mass, and, remarkably, approximately one-sixth had mergers with satellites of mass exceeding 10 times of the host's stellar mass.  These mergers lead to a sharp, but often temporary, decrease in the D/T of the hosts, implying that discs are usually destroyed and then quickly re-grow. We estimate that about half of galaxies today have discs that have grown after massive mergers since $z=2$ and about a third are discs formed post-merger since $z=1$. To form these discs, high cold gas fractions are required post-merger, as well as a relatively quiescent recent history (over a few Gyrs before $z=0$). Our results show quantitatively that discs can form via  diverse merger pathways and that quiescent histories are not the dominant mode of disc formation. The fact that many of these formation pathways lead to systems with similar D/T at the present day, suggests that the variations in the cold gas fraction and merger histories introduce degeneracies that may hamper the reconstruction of merger histories of disc galaxies from observations.
\end{abstract}

\begin{keywords}
Galaxies: formation -- Galaxies: evolution -- Galaxy: disk  
\end{keywords}
\end{frontmatter}

\section{INTRODUCTION }
\label{sec:intro}

A long-standing problem in galaxy formation has been reconciling the abundance of disc galaxies observed in the nearby Universe (e.g. \citealt{weinmann06,park07}) with the high rate of supposedly disc-destroying mergers predicted in $\Lambda$CDM cosmologies \citep{toth92,stewart08,boylan-kolchin10}. For example, the majority ($\sim 70\%$) of Milky Way-mass haloes are predicted to undergo mergers with total mass ratios $M_{\rm sat}:M_{\rm host}>1:10$ \citep{stewart08}, where the total masses include baryons and dark matter.  Although considered `minor' from a total mass ratio perspective, a subhalo with a tenth the mass of the Milky Way's total halo mass has a mass that is roughly twice the mass of the present-day stellar disc.  From the central galaxy's perspective, therefore, such mergers are clearly significant. Combining the merger rates for systems in the $1:10$ mass range with collisionless estimates of merger-induced disc heating rates, one may conclude that very few discs should survive down to the present day.  This is in evident contradiction with local observations, which show that $\sim70\%$ of normal galaxies with stellar masses $\sim10^{10-10.5} M_\odot$ have significant disc components (e.g., \citealt{park07,kelvin14}).

The solution to this problem is to discard the paradigm of quiescent formation history of disc galaxies, which is now generally agreed to be obsolete. Gas physics is commonly invoked to reconcile present-day disc morphologies with active merger histories. Gas-dynamical simulations suggest an interplay of various mechanisms such as feedback and gas-rich mergers \citep{barnes02,springel03,springel05a,robertson06,hopkins09,scannapieco09,brooks11,brook12}, accretion of gas through cold flows (e.g., \citealt{governato09}) and/or cooling of the hot phase (e.g., \citealt{moster12}).  The presence of cold gas prior to the merger can act to absorb (and later radiate) some of the unbinding energy injected during the merger, while any cold gas remaining after the merger, plus any new cold gas that is accreted post-merger, can lead to re-growth of the disc.  

Still needed is a more quantitative assessment of the role of gas in alleviating the effect of mergers. Using controlled and idealised hydrodynamical simulations of galaxy collisions, \citet{hopkins09} found that the survivability of discs was strongly correlated with the initial gas fraction (defined as $f_{\rm gas, b} \equiv {\rm M}_{\rm gas}/{\rm M}_{\rm (gas+stars)}$).   Results of controlled hydrodynamical simulations, where the gas fraction is treated as a free parameter, show that galaxy discs usually re-form when $f_{\rm gas,b} \geq 80-90\%$ (e.g. \citealt{robertson06,hopkins09}). These values appear to be in reasonable agreement with observational gas fraction estimates at $z=2$ \citep{erb06}.  Note, however, that at redshifts $1< z <2$ (where the merger activity is still significant), the cold gas fractions can be as low as $50\%$ (e.g., \citealt{tacconi13}).  Semi-analytical models (e.g. \citealt{delucia04,bower06,font08,somerville08}) obtain similar results, although the results are known to depend on the detailed implementation of physical prescriptions \citep{hopkins10b}.
Since potentially disc-damaging mergers are expected to occur frequently down to $z\simeq 1$ \citep{stewart08,boylan-kolchin10}, it is important to evaluate in a self-consistent way the relation between cold gas fractions, disc formation and the merger history of galaxies.  This can be achieved with cosmological hydrodynamical simulations.

To date, studies based on zoomed cosmological simulations (typically of small numbers of systems) have shown mixed results, at least in terms of the interpretation of how discs persist to the present day.  Some studies support the scenario in which discs form as a result of gas-rich ($f_{\rm gas,b}>0.8$) mergers \citep{brook07,guedes11}, although, in these simulations, the major mergers are restricted to occur at high redshift.  Other studies show that discs can form at low redshift when gas fractions are much lower. For example, \citet{governato09} found a large disc growing after a major (1:1) merger at $z \sim 0.9$, when  $f_{\rm gas,b}<0.25$.  They argue that in this case an important role is played by cold flow accretion along filaments \citep{keres05,keres09,dekel09}, a process which is not included in idealised galaxy merger simulations.  A disc galaxy is also shown to form after $z<1$ in a major merger in the simulation of \cite{brook12}, although this system has a lower mass than a typical disc galaxy.  

Ideally, one would like to tackle this problem using large, statistically-representative samples of galaxies simulated at high resolution with hydrodynamics and in the full cosmological context.  With the advent of large-volume simulation campaigns such as \gimic\ \citep{crain09} and more recently Illustris \citep{vogelsberger14}, EAGLE \citep{schaye15} and Horizon-AGN \citep{kaviraj16}, this is now a possibility.  A major success of these campaigns is their ability to produce large samples of disc galaxies, at least in qualitative agreement with the observations.  From this point of view, the problem appears to be `solved', in that the ubiquity of disc galaxies is consistent with our current theoretical paradigm for structure formation (i.e., $\Lambda$CDM).  However, it is nevertheless still important to elucidate the physics that allows galaxies to retain (or re-form) their discs; analysis of large statistical samples of simulated galaxies should help clarify the somewhat confused picture that has emerged from studies based on zoomed simulations of small numbers of systems.  Furthermore, it has not yet been demonstrated that the predicted {\it distribution} of morphologies from simulations actually agrees with that of real galaxies in a quantitative sense.  We will make such a comparison here. We will also quantify the fraction of galaxies that (re-)form their discs after mergers versus those that form more quiescently. For those disc galaxies that form post-mergers, we also investigate the various pathways that lead to the formation of their discs.

We use the Galaxies-Intergalactic Medium Interaction Calculations ( \gimic) simulations to explore these topics.  As demonstrated in a number of previous studies, \gimic\ is successful at forming large numbers of disc galaxies that match a broad number of observed scaling relations and galaxy properties. Namely, the simulations produce reasonably realistic $\sim L^*$ disc galaxies that match the Tully-Fisher relation, the rotation curves, sizes and star formation efficiencies of low-mass disc galaxies \citep{mccarthy12b}, the cold gas fraction--stellar mass and gas-phase metallicity--stellar mass relations \citep{derossi15}, the number, alignment and the spatial distribution of their satellite galaxies \citep{deason11,font11}, the surface brightness and metallicity distributions of the stellar haloes in the Milky Way, M31 and other nearby disc galaxies \citep{font11,mccarthy12a} and the X-ray scaling relations for normal disc galaxies \citep{crain10}.  Therefore these simulations are well-suited to address the topics in our study.

A potential drawback of these simulations is that they do not include a physical prescription for AGN feedback, as has been found to be important in more recent simulations such as EAGLE \citep{schaye15,crain15}.  However, this is not a serious omission for the mass range we consider here (stellar masses of $\sim10^{10}$ M$_\odot$).  We demonstrate this later (Appendix \ref{sec:appendix_eagle}), where we compare the stellar mass$-$halo mass relations, size$-$mass relations, and distribution of morphologies predicted by \gimic and EAGLE, showing that they are very similar for all but the most massive galaxies.  The realism of our simulated disc galaxies is tested further by comparing their morphologies with those derived from recent observations (see Section \ref{sec:disc_prop} and Appendix \ref{sec:appendix_gama}).

This paper is organised as follows. In Section \ref{sec:sims_methods} we summarise the simulations and sample selection.  In Section \ref{sec:disc_prop} we compare the morphologies of the simulated galaxy population with observational data.  In Section \ref{sec:probs} we calculate halo--halo and subhalo-galaxy merger rates and show the effects of mergers of different mass ratios on the simulated discs.  In Section \ref{sec:coldgas_and_mergers} we investigate the role played in the growth of stellar discs by cold gas fractions and merger histories and quantify the various formation histories of present-day disc galaxies. In Section \ref{sec:imprints} we discuss the disc re-growth scenario in the context of some observations.  Finally, the main conclusions are summarised in Section \ref{sec:concl}. 

\section{Simulations and Sample Properties}
\label{sec:sims_methods}

\subsection{Simulation description}
\label{sec:sims}


In this study we use the two highest resolution simulations from the \gimic suite of simulation  (see \citealt{mccarthy12b}). Following the \gimic terminology \citep{crain09}, these regions are called \smm and \sz, which indicates that the overdensities of the two regions at $z=1.5$ are $-2$ and $0$, respectively, of the standard deviation from the cosmic mean $\sigma$ (here $\sigma$ is the \textsc{RMS} mass fluctuation on a scale of $18$~h$^{-1}$ Mpc). The two roughly spherical regions, with radii of $\sim 18 h^{-1}$Mpc, were extracted from the dark matter-only Millennium simulation \citep{springel05b} and re-simulated from $z=127$ to $z=0$ with gas dynamics. The remaining Millennium volume ($500$ $h^{-1}$ Mpc on a side) was also simulated but at lower resolution and with purely collisionless dynamics. The cosmological parameters used in these runs are: $\Omega_{\rm m}=0.25$, $\Omega_{\rm \Lambda}=0.75$, $\Omega_{\rm b}=0.045$, $\textrm{n}_{\rm s}=1$ (the spectral index of the primordial power spectrum), $\sigma_{8}=0.9$ (the RMS amplitude of linear mass fluctuations on a $8$~$h^{-1}$ Mpc scale at $z=0$), $\textrm{H}_{0}=100~h~\textrm{km s}^{-1}\textrm{Mpc}^{-1}$ and $h=0.73$.

The increased resolution enables us to better resolve the structure of galaxy discs and the high-mass end of substructure mass function, $\textrm{M}_{\rm sat} \geq 10^{8-9}~\textrm{M}_{\odot}$. The masses of dark matter particles are $6.63\times10^{6}~h^{-1}~\textrm{M}_{\odot}$, those of gas particles are $1.46\times10^{6}~h^{-1}~\textrm{M}_{\odot}$, and the gravitational softening is $0.5 \, h^{-1}$~kpc.  This is comparable with the resolutions of the recent Illustris \citep{vogelsberger14} and EAGLE (\citealt{schaye15,crain15}) main box simulations.
The dark matter resolution is similar to other large-scale (but dark matter only) cosmological simulations, such as Millennium II \citep{boylan-kolchin09}, which enables us to perform a comparison with their results.

The simulations were run with the TreePM-SPH code \gadget\ (last described by \citealt{springel05b}).  Below we only summarise the subgrid physics prescriptions in this code and direct the reader to the referenced papers for further details.  The code includes prescriptions for star formation \citep{schaye08}, metal-dependent radiative cooling in the presence of a global UV/X-ray background \citep{wiersma09a}, stellar mass loss and chemical evolution \citep{wiersma09b} and a kinetic supernova feedback model \citep{dallavecchia08}.  

The simulations lack the resolution to resolve the Jeans scales of the cold ($T < 10^4$ K), dense gas phase ($n_{\rm H} \gg 0.1$ cm$^{-3}$) of the ISM.  An effective equation of state is instead imposed, in order to approximate the effects of feedback and turbulence in the ISM and to prevent artificial fragmentation for gas that is above a fixed density threshold of $n_{\rm H} > 0.1$ cm$^{-3}$.  Gas particles on the equation of state are allowed to convert into star particles (stochastically) at a pressure-dependent rate that reproduces the observed Kennicutt-Schmidt law \citep{kennicutt98}, by construction.  Radiative cooling is computed on an element-by-element basis, following 11 species relevant to cooling.  Feedback associated with star formation is modelled using a kinetic implementation (i.e., neighboring gas particles are given a velocity kick), with an initial wind velocity of $600$ km s$^{-1}$ and a mass-loading parameter (i.e., the ratio of the mass of gas given a velocity kick to that turned into newly formed star particles),  $\eta$, set to $4$.  This corresponds to using  approximately 80\% of the total energy available from supernovae for a \citet{chabrier03} IMF, which is assumed in the simulation.  This choice of parameters results in a good match to the peak of the star formation rate history of the Universe \citep{crain09,schaye10}.

As described in the Introduction, these simulations have achieved a number of notable successes.  At the same time, it is important to acknowledge their shortcomings.  Most notably, because they neglect feedback from AGN, the simulations suffer from overcooling in the most massive galaxies.  As shown by \citet{crain09}, the simulations fail to reproduce the bright end of the galaxy stellar mass function, in that they predict too many massive galaxies compared to what is observed.  However, in the present study, we focus on galaxies with $z=0$ stellar masses of M$_{\rm star} \sim10^{10}$ M$_{\odot}$, which is the typical mass of normal disc galaxies today \citep{kelvin14}.  The simulations also reproduce many of the observed correlations of lower mass galaxies (see \citealt{mccarthy12b}), but we focus on systems with M$_{\rm star} \sim10^{10}$ M$_{\odot}$ as they are the best resolved disc galaxies in our simulations.  We reiterate that in Appendix \ref{sec:appendix_eagle} we show the galaxies used in this study agree on a number of key relations and properties with galaxies of similar mass in the EAGLE simulations (which better reproduce the full galaxy stellar mass function), indicating that AGN feedback does not play an important role for the galaxies under consideration.

\subsection{Identifying subhaloes and constructing merger histories}
\label{sec:subhaloes}

Haloes are identified using a standard friends-of-friends (FoF) algorithm with a linking length of $b=0.2$ times the mean inter-particle separation, run on the dark matter particles.  Baryons are included by linking in the nearest baryon particle (gas or star) to each FoF dark matter particle.  All groups with a minimum of 20 particles are retained for further analysis.  We then remove all haloes that are not gravitationally self-bound by performing an unbinding calculation with \subfind \citep{springel01,dolag09}. 

Simple merger trees are constructed for the main haloes that satisfy our mass selection criterion (see Section \ref{sec:MWs} below).  We trace the main progenitors of the $z=0$ systems back to $z\simeq 5$, before the emergence of any discs. The procedure for tracing back the main progenitor of the central halo is as follows. We select all dark matter particles within $r_{200}$ of the main halo at $z=0$ and identify them at earlier redshifts using their unique IDs. At each redshift output, we determine whether these particles are bound gravitationally to any structure. The main progenitor at a given redshift is taken to be the subhalo that contains the largest fraction of the dark matter particles selected at $z=0$.  The algorithm uses only the dark matter particles, as they constitute most of the mass of a system and, unlike baryonic particles, are not directly affected by pressure forces.  Having identified the main progenitor, we then follow forward in time from $z\simeq 5$ all (sub)haloes that eventually merge with the main progenitor's FoF group by the present day.

\subsection{Galaxy sample}
\label{sec:MWs}

We kinematically classify the galaxies into disc- and spheroid-dominated categories based on the rotational angular momentum, $J_z$, of bound star particles within 20~kpc (we use the most bound particle to identify the galaxy center).  To assign particles to the disc component we use a simple cut of $J_z/J_{\rm circ} \ge 0.75$, where $J_{\rm circ}$ is the angular momentum of a star particle on a co-rotating circular orbit with the same energy.  In our definition, disc galaxies are those with a kinematical disc to total ratio $\textrm{D}/\textrm{T} \ge 0.3$, however the main conclusions of this paper do not change by choosing a higher limit of $\textrm{D}/\textrm{T}$ or by adopting a different cut in the angular momentum required for disc assignment.  In Section \ref{sec:disc_prop} we discuss the correspondence between the 3D kinematic D/T diagnostic that we employ throughout the present study and the apparent (2D) morphology that one would more typically derive for observed galaxies.

As in our previous studies \citep{font11,mccarthy12a,mccarthy12b}, we do not impose any constraints on the merger histories of galaxies, as we also want to follow the formation of disc galaxies with non-quiescent merger histories (in addition to galaxies that contain no significant disc component, to investigate why).  We select only galaxies in the range of total stellar mass $9.7 \leq $ log(M$_{\rm star, \rm tot}$/M$_{\odot}) \leq 10.3$ (median stellar mass of $\simeq 10^{9.9}$ M$_{\odot}$).  This mass range is chosen in order to minimise the mass-dependence in the merger rates results and also to avoid issues with overcooling at higher masses (see \citealt{mccarthy12b}).  For reference, the upper M$_{\rm star, \rm tot}$ limit is somewhat smaller than the estimated stellar mass of the Milky Way, M$_{\rm star, \rm tot,{\rm MW}} \simeq 3.6-5.4 \times 10^{10}$ M$_{\odot}$ \citep{flynn06,mcmillan11}.  The median halo mass M$_{200}(z=0)$ (i.e. the mass enclosed within a sphere containing an average density $\simeq 200$ times the present-day critical density) is $\simeq 10^{11.7}$ M$_{\odot}$, is also somewhat lower than the typical $1-2 \times 10^{12}$ M$_{\odot}$ inferred for the Milky Way \citep{evans00,battaglia05,karachentsev06,li08,guo10}, although some recent studies estimate a lower total mass of the Milky Way, $\sim 5-10 \times 10^{11}$ M$_{\odot}$ \citep{xue08,deason12,gibbons14}.  Therefore, our simulated galaxy sample is more representative of slightly sub-L$^*$ galaxies. Note that even within this mass range we still have a large sample of 107 high-resolution simulated galaxies. As already mentioned, this mass range also encompasses the typical stellar mass of disc galaxies observed in the local Universe \citep{kelvin14}.

Finally, we note that the sample combines systems from both the \smm and \sz regions since galaxy properties do not show any significant dependency on very large-scale environment (see also \citealt{crain09,font11}).

\section{Present-day morphologies: comparison with observations}
\label{sec:disc_prop}

Before proceeding to our main analysis, we first examine some additional relevant tests of the realism of the present-day discs in the simulations (as mentioned in the Introduction, the simulated galaxies match a broad range of global scaling properties of present-day galaxies).  Specifically, here we focus on the present-day morphologies of the simulated galaxies and compare them, in a like-with-like fashion, to observational data, demonstrating that the simulated galaxy population has a realistic distribution of morphologies.

We first compare to the observed morphologies of local galaxies in the GAMA survey.  From the GAMA Data Release 2 database\footnote{http://www.gama-survey.org/dr2/}, we select local galaxies with $0.07 < z < 0.12$ (median $z\approx0.10$) and that lie in the same stellar mass range as the simulated galaxies (i.e., $9.7 < \log_{10}[M_*/{\rm M}_\odot] < 10.3$) and that have had single S{\'e}rsic models fitted to their surface brightness profiles in the various GAMA bands by \citet{kelvin12}.  Applying these selection criteria yields a sample of approximately 4,800 galaxies. We focus on the S{\'e}rsic index, which governs the rate of the fall off of light with radius and therefore characterises the morphology of a galaxy. More details about the construction of synthetic GAMA r-band images and about the S{\'e}rsic index model fitting are given in Appendix \ref{sec:appendix_gama}.

\begin{figure}
\includegraphics[width=0.995\columnwidth]{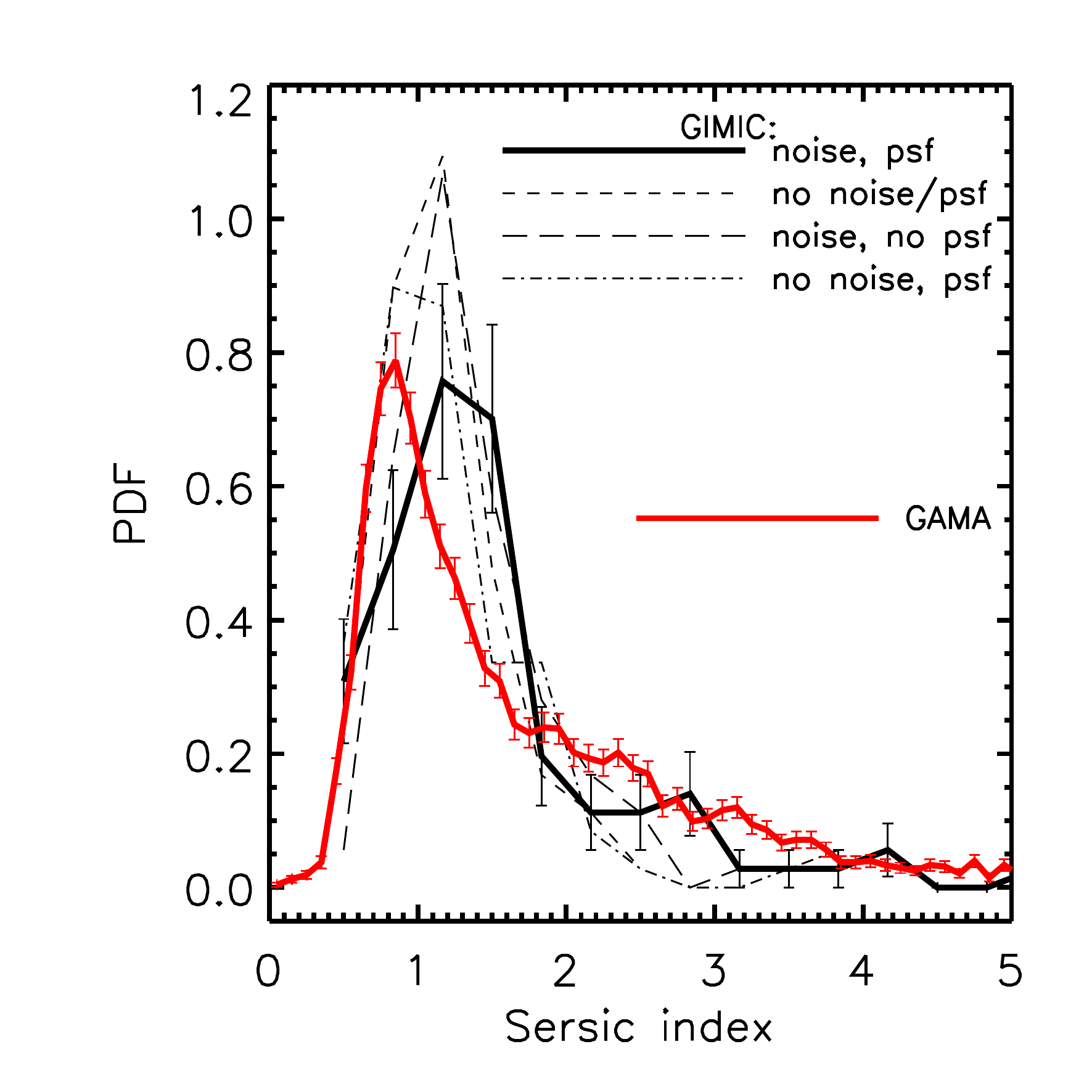} 
\caption{Comparison of the distribution of S{\'e}rsic indices from \gimic\ (thick solid black curve) with that derived from the analysis of local ($0.07<z<0.12$) GAMA galaxies, selected to be in the same stellar mass range applied to the simulated galaxies (thick solid red curve; \citealt{kelvin12}).  The error bars correspond to the Poisson errors, derived by taking the square root of the number of galaxies in each S{\'e}rsic index bin.  To derive the S{\'e}rsic indices of the simulated galaxies, we create synthetic GAMA-like images of the simulated galaxies, accounting for the effects of the SDSS telescope point spread function and Poisson noise (due to both the galaxy and the sky; see Appendix \ref{sec:appendix_gama} for further details).  The thin short-dashed, long-dashed, and dot-dashed black curves show the effects of switching on/off the modelling of Poisson noise and the point spread function.  The simulated galaxy population has a qualitatively similar distribution of morphologies to that of the observed GAMA sample.
}
\label{fig:gama}
\end{figure}

In Fig.~\ref{fig:gama} we compare the predicted (solid thick black curve) and observed (solid thick red curve) distributions of the r-band S{\'e}rsic index. Overall, the simulated distribution is quite similar to that of the GAMA sample, both showing strong peaks in the distribution near a S{\'e}rsic index of 1 (which corresponds to an exponential disc profile).  In detail, the peak of simulated distribution occurs at a slightly higher value than that of the GAMA sample.  Nevertheless, the agreement is impressive considering that no attempt was made to calibrate the \gimic\ simulations on any aspect of the morphology of the galaxies.  

The thin black curves show the effects of turning on/off the modelling of the noise and the point spread function.  Including these effects does not change the qualitative picture, but it does affect the derived distribution in a quantitative sense, in that the resulting distribution is somewhat broader and the peak is shifted to slightly larger values of the S{\'e}rsic index when realistic noise and smoothing is incorporated into the analysis.

\begin{figure}
\includegraphics[width=0.995\columnwidth]{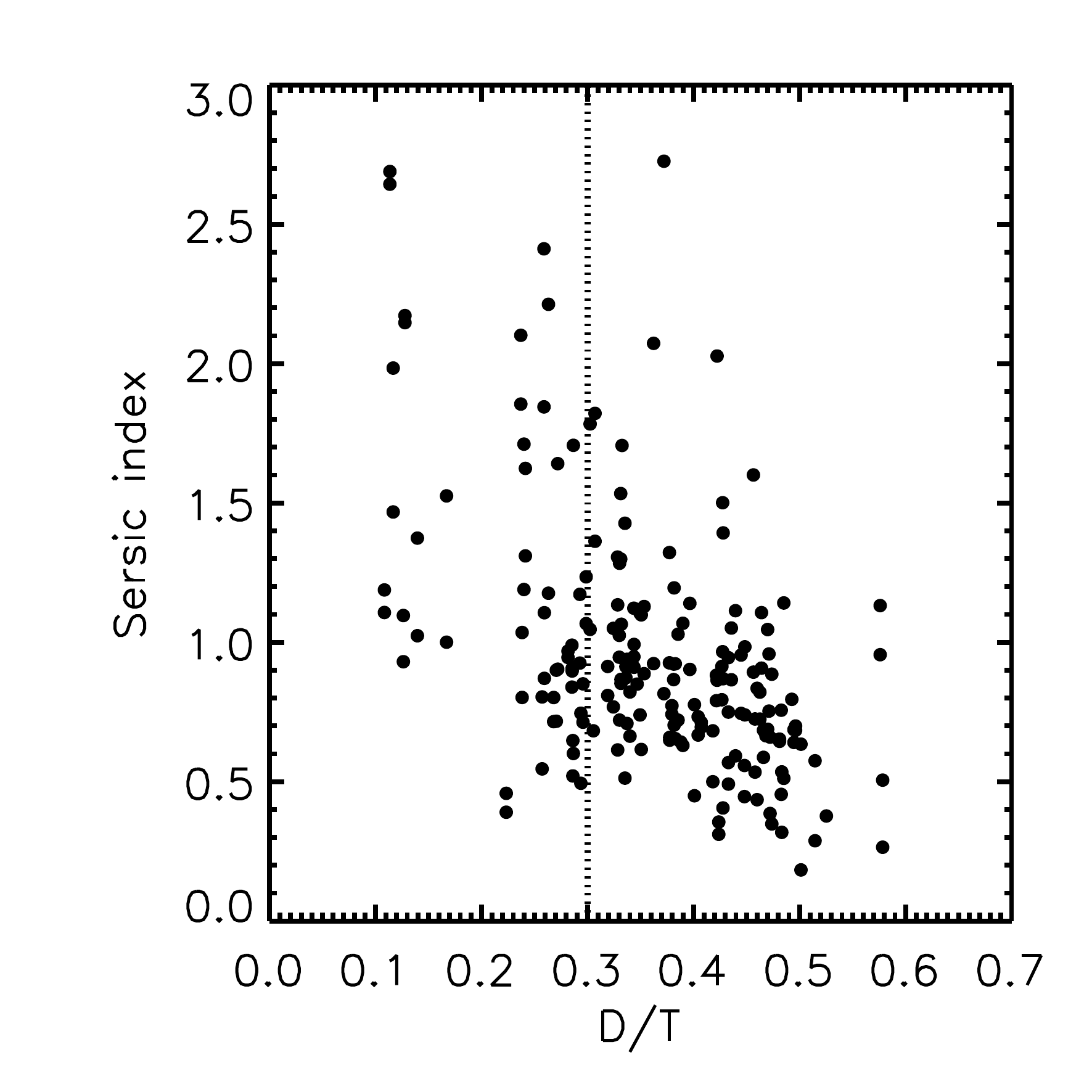} 
\caption{The $z=0$ relation between the kinematic D/T and the apparent (2D) morphology, as characterised by the S{\'e}rsic index derived from synthetic images of \gimic\ galaxies.  Although significant scatter is present, a strong anti-correlation is clearly visible, such that high values of $D/T$ correspond to low values of the S{\'e}rsic index.  The vertical dotted line at D/T=0.3 corresponds to our fiducial dividing line between disc-dominated and bulge-dominated systems, which also roughly delineates the systems into high and low values of the S{\'e}rsic index.}
\label{fig:n_DT}
\end{figure}

It is of interest to examine briefly the correspondence between the kinematic D/T morphological indicator that we use for the remainder of the paper with the apparent (2D) morphology, as characterised by the S{\'e}rsic index derived from the image modelling analysis above.  In Fig.~\ref{fig:n_DT} we examine the relation between these two quantities at $z=0$.  Although there is significant scatter, there is also clearly a strong anti-correlation between the parameters (the Spearman rank correlation coefficient is $r=-0.49$).  Our fiducial threshold of D/T=0.3 used to separate galaxies into disc- or spheroid-dominated systems also roughly delineates systems into (relatively speaking) low and high values of the S{\'e}rsic index.  Thus, in a broad sense, there is a good correlation between the 3D kinematics and the 2D (light-weighted) appearance of galaxies, which is expected and reassuring (see also \citealt{scannapieco10,obreja16}).  We note that we use the (3D) kinematic D/T as our indicator of the morphology because it is a more physical quantity linked to the energetics of the system and does not depend on complicating factors such as projection effects and the mapping between light and mass (which depends on age, metallicity, etc.).

As shown above, many of the simulated and observed galaxies in this stellar mass range have a broadly disky appearance, in that they are typically characterised by S{\'e}rsic indices of $\sim1$.  One can go a step further and ask whether the detailed distribution of scale-heights ($z_{0}$) and scale-lengths ($h_{R}$) of (thick+thin) components of the simulated discs are also reasonable.  For this test, we compare with measurements of these quantities of similar-mass nearby late-type galaxies from \citet{kregel02} and \citet{yoachim06}.

For a given simulated galaxy we compute the scale-height and -length in the following manner (which is meant to roughly mimic the observational analysis).  We first rotate the galaxy into an edge-on configuration, by aligning the total angular momentum of the stars (within a 20 kpc aperture) with the z-axis of the simulation box.  We then select the disc particles and produce two-dimensional i-band surface brightness maps in cylindrical coordinates $(R,z)$, with $R$ being the projected radius along the disc.  Following the observational studies that we compare to, we fit a two-dimensional parametric model of the following form to the maps:
\begin{equation}
L(R,z)=L_0\exp^{-R/h_R}f(z) \ \ \, 
\end{equation}

\noindent where $L_0$ is the central luminosity density and $f(z)$ is the generalized $sech^{2/N}(Nz/z_0)$ vertical distribution ($N=1$ for the thin+thick disc; see \citealt{yoachim06}).  

\begin{figure*}
\includegraphics[width=0.995\columnwidth]{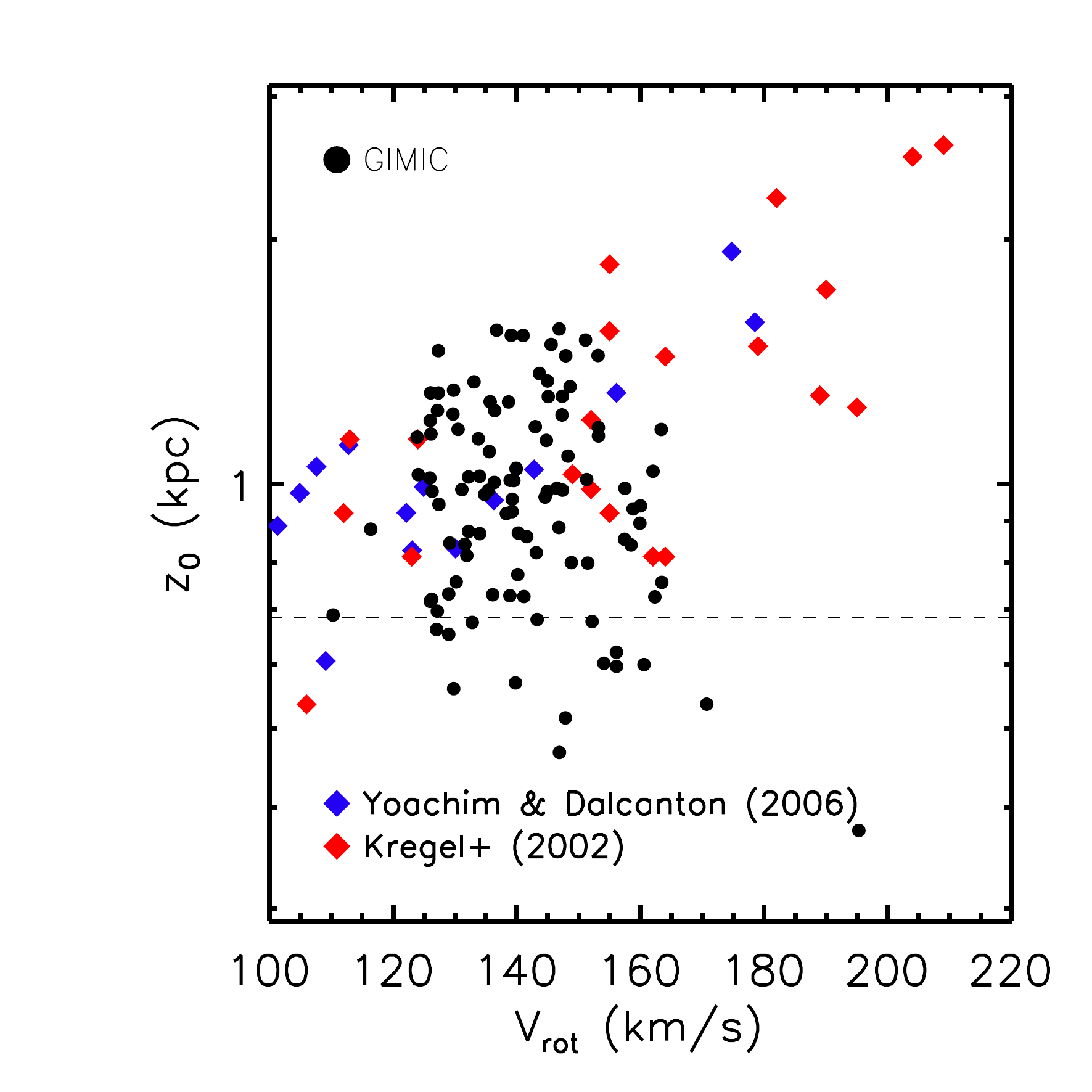} 
\includegraphics[width=0.995\columnwidth]{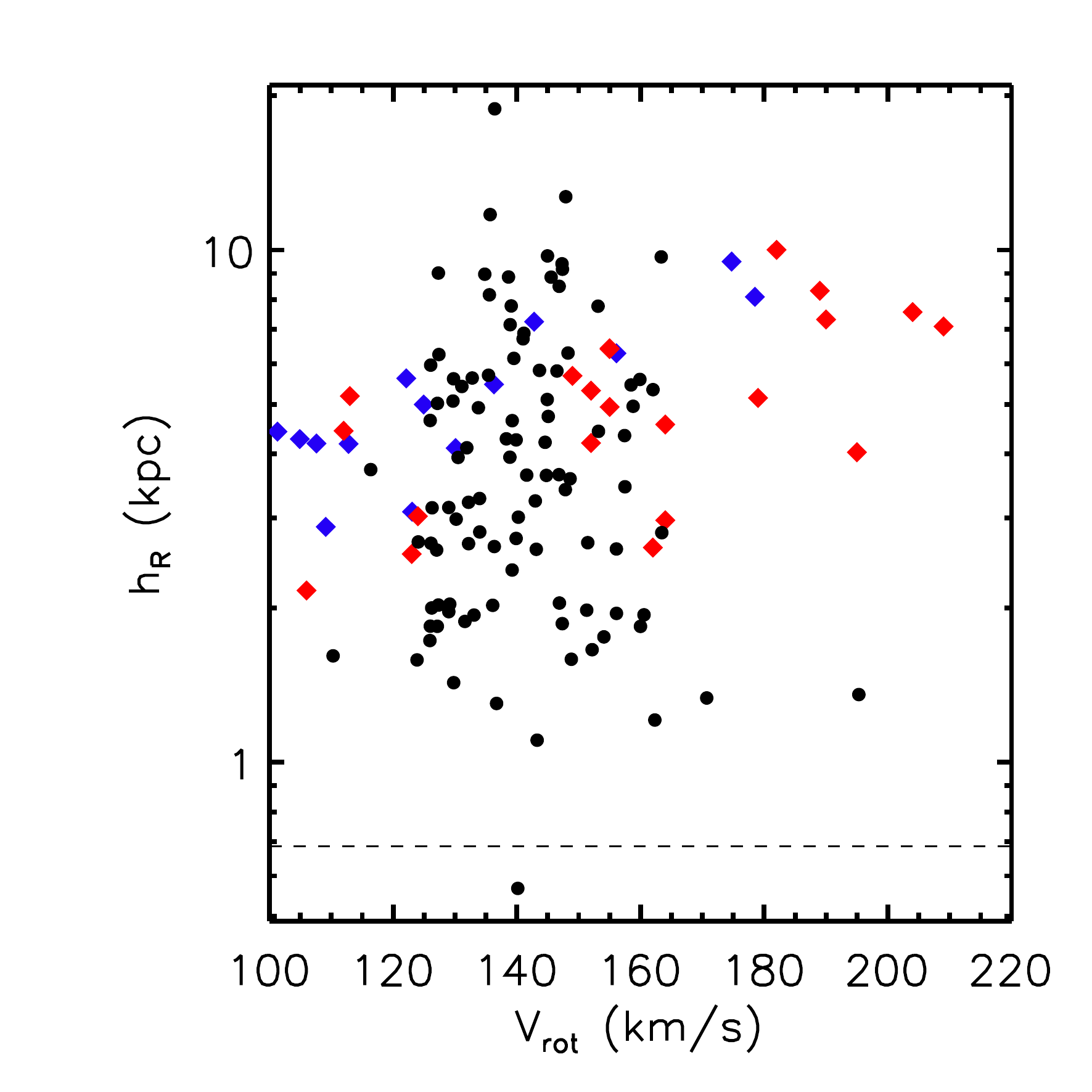}
\caption{Comparison of disc scale-heights ({\it Left}) and scale-lengths ({\it Right}) of simulated galaxies in \gimic (black circles) with those measured for nearby galaxies by \citet{kregel02} and \citet{yoachim06} (red and blue diamonds) versus the maximum rotation speeds $V_{\rm rot}$.  The dashed horizontal line shows the (Plummer equivalent) force softening of the simulations.  Over the range of masses considered here, the simulated galaxies have broadly realistic sizes.}
\label{fig:disc_scales}
\end{figure*}

Fig.~\ref{fig:disc_scales} shows the present-day disc scale-heights and scale-lengths versus the maximum rotation speeds $V_{\rm rot}$ of the sample disc galaxies, in comparison with similar measurements in nearby sub-L$^*$ disc galaxies of \citet{kregel02} and \citealt{yoachim06}. The typical values for the simulations of $z_{0} \sim 1$~kpc and $h_{R} \sim 4$~kpc are in reasonably good agreement with the observational data (see also the data of \citealt{bizyaev02}).  However, we caution that the comparison to the scale-heights may be somewhat influenced by the force resolution of the simulation (the Plummer-equivalent softening is $500 h^{-1}$ pc, represented by the dashed line in Fig.~\ref{fig:disc_scales}) and the simplified treatment of the ISM (the Jeans length of the equation of state is $\sim 1$ kpc), which will effectively prevent the formation of very thin discs in the simulations.  Nevertheless, it is encouraging that the structural properties are similar to observed nearby galaxies, which gives us some confidence that the disc component in the simulated galaxies will react to mergers, etc. in a realistic fashion.


\subsection{Summary of comparison to observations}

We have shown that the simulated galaxy population that we examine in this study has a reasonably realistic distribution of morphologies, as characterised by the S{\'e}rsic index and effective radii and the scale-lengths and scale-heights of the disc component.  The sample is therefore well-suited to address the question of the origin of present-day morphologies and elucidating its connection to the assembly history of galaxies.

\section{Mergers: probabilities and effects}
\label{sec:probs}

In this section we revisit the likelihood of mergers of different mass ratios in a $\Lambda$CDM cosmology.  We demonstrate that most $L_{*}$ galaxies (${\rm M}_{\rm star}\sim10^{10} M_\odot$) are expected to experience a collision with mass ratio $\sim1:1$ (defined in terms of the ratio of total satellite mass to central galaxy stellar mass) since $z\sim2$.  We further show that such mergers are generally expected to induce large morphological transformations.

\subsection{Merger probabilities}
\subsubsection{By the ratio of total halo masses}
 
Several studies have already calculated the merger rates of Milky Way-mass dark matter haloes using N-body cosmological simulations (e.g. \citealt{stewart08} and \citealt{boylan-kolchin10}). In particular, the Millennium II simulations used by \citet{boylan-kolchin10} have many similarities with \gimic, for example the identical cosmologies and the similar mass resolutions. Although smaller in volume, our gas-dynamical simulations take into account additional mechanisms associated with the existence of stellar discs. We first briefly examine halo--halo mergers to provide a baseline for comparison with dark matter only studies.

\begin{figure}
\includegraphics[width=0.995\columnwidth]{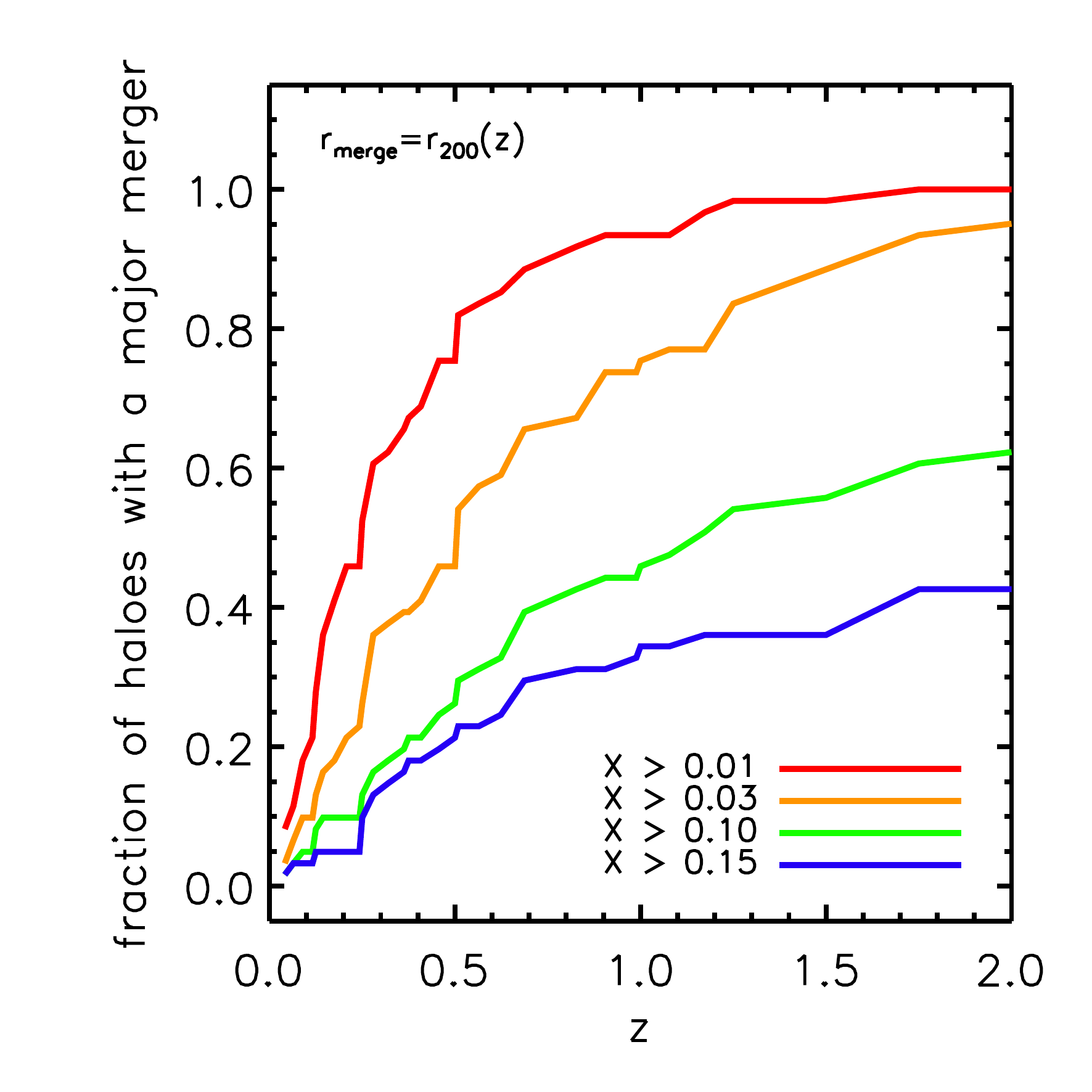}
\caption{The cumulative fraction of haloes (median M$_{200}=10^{11.7}$ M$_{\odot}$) which have mergers with mass ratios $X\equiv\textrm{M}_{\rm sat, max}/\textrm{M}_{200}(z=0) > 0.01$ (red), $>0.03$ (orange), $>0.10$ (green) and $>0.15$ (blue).}
\label{fig:halo_merger_frac}
\end{figure}

Fig.~\ref{fig:halo_merger_frac} shows the cumulative fraction, $f$, of all the main dark matter haloes in our sample that undergo mergers with mass ratios $X \equiv M_{\rm sat, max}/{\rm M}_{\rm 200} (z=0)$ (where M$_{\rm sat, max}$ is the maximum mass of a dark matter subhalo over its whole history and M$_{\rm 200}$ is the virial mass of the host halo at $z=0$) greater than several threshold values.  The results are quite similar to those of \citealt{boylan-kolchin10}, who find $\simeq 55\%$ of haloes had mergers $> 1 : 10 $ and $> 30\%$ had mergers $> 1 : 6-7$.  As already noted, the masses of these merging systems are $2-3$ times more massive than the mass of a putative Milky Way disc embedded in these dark matter haloes.  Therefore, on the basis of halo--halo merger probabilities, one may conclude that a significant fraction of dark matter haloes experience mergers which are potentially damaging to the embedded discs from $z=2$ to the present.  Moreover, some haloes have multiple events of this kind \citep{stewart08,boylan-kolchin10}.

\subsubsection{By the ratio of satellite to central galaxy mass}

\begin{figure}
\includegraphics[width=0.995\columnwidth]{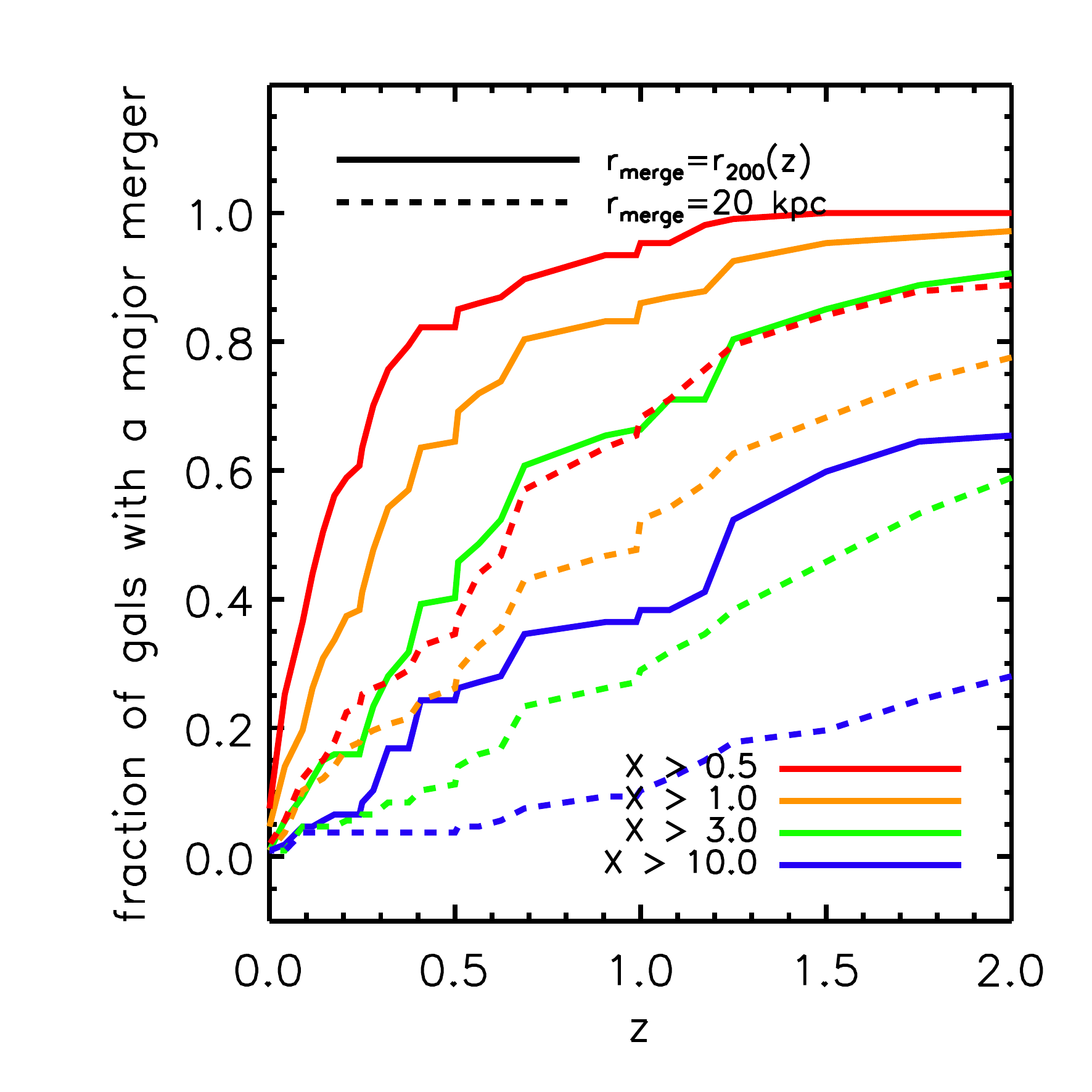}
\caption{The cumulative fraction of galaxies that have mergers with mass ratios: $X\equiv\textrm{M}_{\rm sat}/\textrm{M}_{\rm star, host} > 10$ (blue), $>3$ (green), $1.0$ (orange), and $>0.5$ (red). Solid lines correspond to the case when the satellite crossed $r_{200}(z)$ and dashed lines when the satellite crossed a fixed physical radius of $20$~kpc.}
\label{fig:merger_frac}
\end{figure}

The merger probabilities involving (sub)haloes and the central galaxy are more directly relevant to the question of the survivability of galaxy discs than the halo-halo merger rates. Fig.~\ref{fig:merger_frac} shows the cumulative fraction of galaxies that undergo mergers with mass ratios $X\equiv{\rm M}_{\rm sat}/{\rm M}_{\rm star, host} >0.5,1,3,10$.  We explore the importance of mass loss of the satellite due to {\it tidal stripping} since the time of accretion onto the main halo, by computing the mass ratio both at the time of accretion (defined as when the satellite first crosses $r_{200}(z)$ of the main halo) and when it first comes within 20 kpc (physical) of the central galaxy (i.e., roughly the size of a typical galaxy disc, around which tidal interactions are the strongest). These two cases are represented by the solid and dashed curves, respectively.  

When using the mass ratio at the time of accretion to calculate the probabilities, it is clear that major mergers are quite common.  For example, nearly all galaxies are expected to have had at least one merger where the incoming satellite is of comparable mass (1:1) to the central galaxy (in terms of its stellar mass).  However, the picture changes significantly when one defines the merger mass ratio using the satellite mass just prior to the collision (here evaluated at 20 kpc).  For example, when tidal stripping is taken into account, $\sim 20-30\%$ fewer galaxies experience minor mergers ($X>0.5-1$) since $z=2$ and significantly fewer galaxies, about a factor of three fewer, experience mergers with $X>10$. The sharp decrease in the galaxy merger probabilities with increasing mass ratio $X$ illustrates that more massive satellites are more strongly affected by tidal stripping.  Overall, our results suggest that semi-analytic models that neglect this effect will markedly overestimate the merger rates of galaxies and possibly also the damage induced to the embedded discs (see also \citealt{wilman13}).

Nevertheless, tidal stripping by itself is not a solution to the disc abundance problem, since the rate of disc-damaging mergers is still high even after its effect is taken into account. For example, the dashed lines in Fig.~\ref{fig:merger_frac} show that since $z=2$ the majority ($\sim 60\%$) of galaxies in the sample undergo mergers $>3 {\rm M}_{\rm star, host}$ and $\sim 80\%$ undergo mergers $> {\rm M}_{\rm star, host}$.  As we show below, most of these mergers are expected to induce significant morphological changes to their hosts. Therefore disc-damaging mergers are still expected to be frequent in a hierarchical scenario, in qualitative agreement with previous dark matter only results.  

\subsection{The effect of mergers on galaxy discs}

\begin{figure}
\includegraphics[width=0.995\columnwidth]{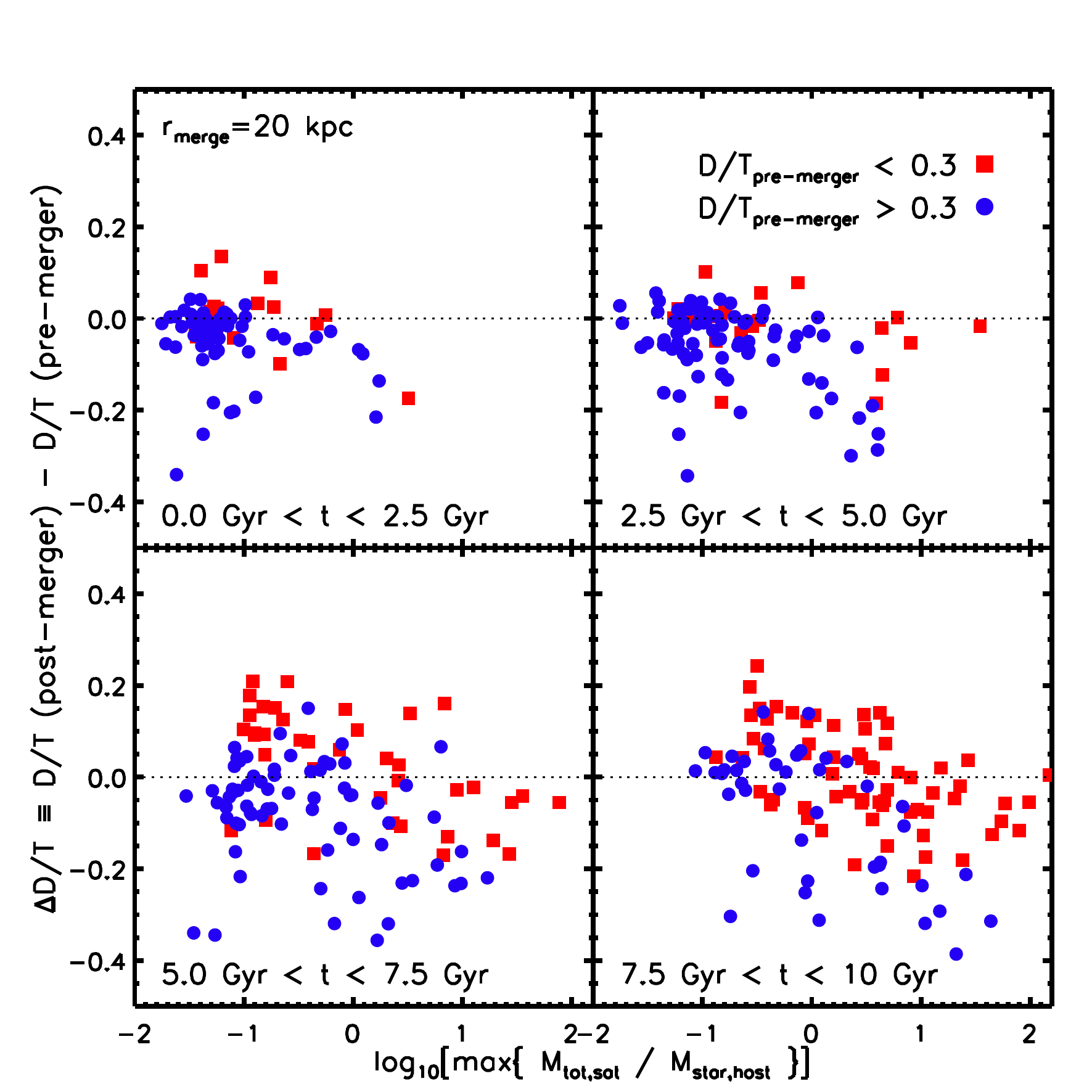}
\caption{Changes in D/T in host galaxies measured just after and before the merging of the most massive satellite (taken when the satellite crosses at $r_{\rm merge}=20$~kpc). The panels show  mergers in different time intervals, from the most recent (top left) to the earliest (bottom right). Blue circles correspond to $\Delta$(D/T) in disc galaxies and red squares to $\Delta$(D/T) in spheroids. Significant D/T changes occur typically for satellites ${\rm M}_{\rm sat} \geq {\rm M}_{\rm star, host}$, and these changes are more pronounced for disc galaxies.}
\label{fig:delta_DT_mass_frac}
\end{figure}

Having calculated the likelihood of mergers of different mass ratios, we now assess the effect that these mergers have on the disc, by evaluating the change they induce in the D/T ratio.  In Fig.~\ref{fig:delta_DT_mass_frac} we plot $\Delta$(D/T) of host galaxies measured between two snapshots on either side of the time of the most massive merger, defining the merger time as that when the satellite crosses a fixed distance of $r_{\rm merge}=20$~kpc.  The four panels show the mergers in different lookback time intervals, from the most recent (top left) to the earliest mergers (bottom right).  We plot them in lookback time intervals since some galaxies had their most massive merger recently and some had them long ago and it is not immediately obvious that high-redshift galaxies should respond to mergers of a given mass ratio in the same way as low-redshift galaxies (e.g., because they have different sizes, gas mass fractions, etc.).  However, inspection of the different panels suggests that, at least roughly, the mass threshold above which satellites are able to induce significant changes is about that of the stellar component of their hosts, ${\rm M}_{\rm sat} \simeq {\rm M}_{\rm star, host},$ with no strong dependence on redshift.  Note that higher frequency of data points in the bottom panels indicates that merger histories were more active before $z\sim1$.  As expected, disc galaxies (blue circles) are more sensitive to morphological changes than spheroids (red squares) and these changes are almost always towards dispersion-supported systems [i.e., $\Delta$(D/T)$<0$].  This corresponds to a positive change in the S{\'e}rsic index, $\Delta$(n)$>0$, as D/T is anti-correlated with $n$ (Fig. \ref{fig:n_DT}).

The large scatter in the $\Delta$(D/T) data points suggests that changes in galaxy morphology depend on other factors as well, for example on the amount of cold gas contained in these galaxies, orbital parameters, etc. In a {\it few} cases, galaxies can even show an increase in $\Delta$(D/T) after mergers [see data points with $\Delta$(D/T)$>0$].  This may suggest that some galaxies become more disc-like as a result of mergers, possibly fueled by the consistent amount of cold gas which accompanies these mergers; e.g., a bulge-dominated host galaxy could accrete a low-mass, gas-rich companion that brings in sufficient gas to build a disc.  Most of the data points with $\Delta$(D/T)$>0$ indeed correspond to galaxies which were spheroidal pre-merger.  Note, however, that in their case, an increase in D/T does not necessarily translate into a lowering of the S{\'e}rsic index sufficiently close to $1$, as can be inferred from the large range of possible D/T - $n$ trajectories in Fig \ref{fig:n_DT}. The role of cold gas will be analysed in more detail in the following sections.

\section{Factors that influence the present-day morphology}
\label{sec:coldgas_and_mergers}

We have shown above that the majority of galaxies in the mass range under consideration undergo mergers that are expected to damage significantly (and possibly disrupt completely) pre-existing stellar discs.  In spite of this, the majority of galaxies that are at present day in this mass range, both in nature and in these simulations, contain significant disc components.  This suggests that discs can be re-established after mergers, via new star formation in gaseous discs (e.g., \citealt{robertson06,hopkins09}).  But not all the present-day simulated systems have significant discs - why not?  And for those that do contain significant disc components, there is a large spread in `diskiness' (e.g., D/T).  What determines this spread?  Clearly, relevant factors, besides the mass ratio of the merger, are going to be the time of the last massive merger (i.e., is there sufficient time between the merger and the present-day to reform a significant disc component?) and the amount of gas available post-merger to fuel the re-growth.  We now examine the importance of these factors.

\subsection{Dependence on merger mass ratio}

\begin{figure}
\includegraphics[width=0.995\columnwidth]{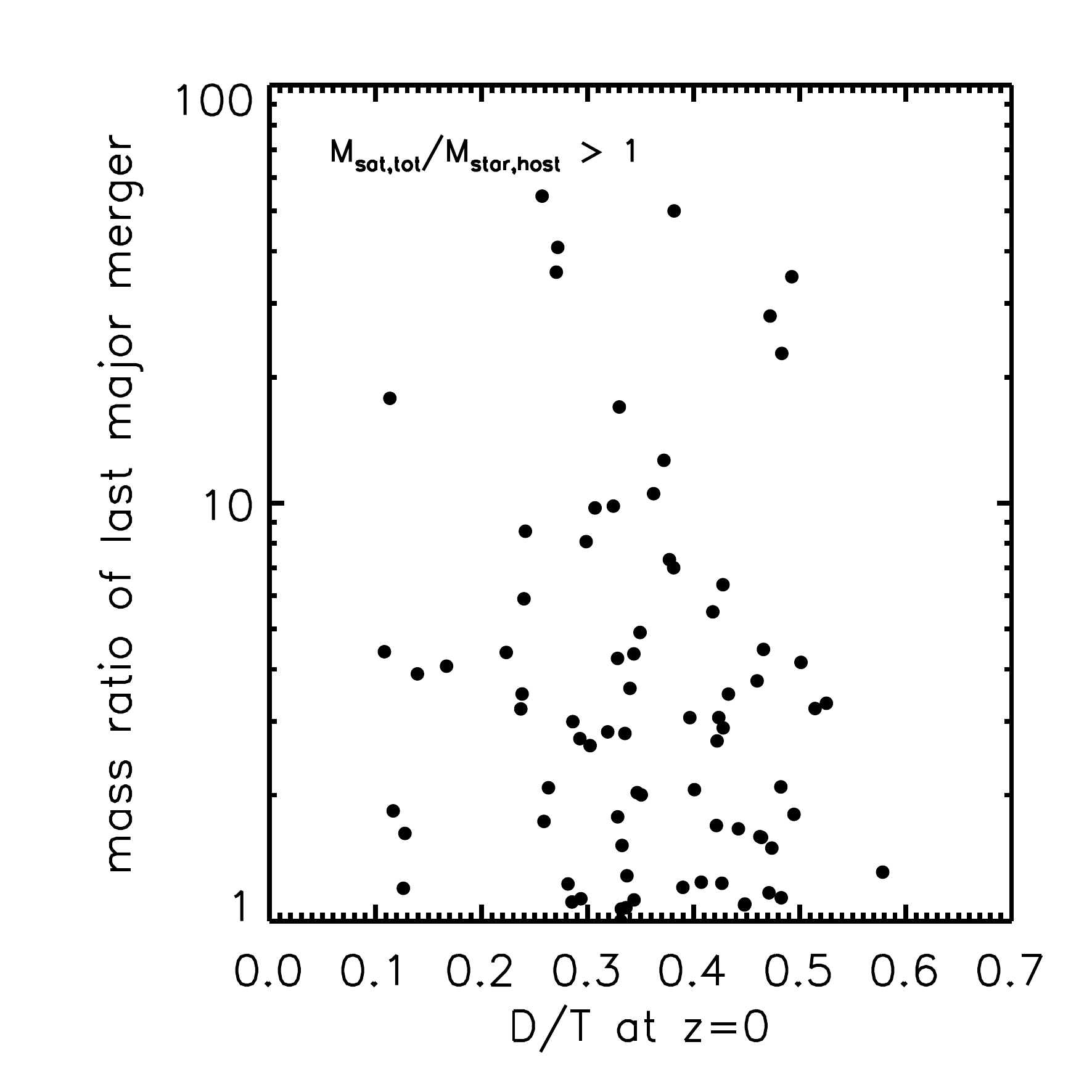}
\caption{Relation between the mass ratio of the last major merger and D/T at z=0.  Even though major mergers do have an immediate effect on the morphology of a galaxy (see Fig.~\ref{fig:delta_DT_mass_frac}), there is virtually no correlation between the present-day morphology and the mass ratio of the last major merger since $z=2$ (Spearman rank correlation coefficient $r=-0.13$).  
}
\label{fig:DT_mass_frac}
\end{figure}

In Fig.~\ref{fig:DT_mass_frac} we plot the relation between the present-day morphology and the mass ratio of the last massive merger.  By selection, we consider as massive mergers only those mergers with mass ratios exceeding unity, ${\rm M}_{\rm sat}/{\rm M}_{\rm star, host} >1$ , guided by Fig.~\ref{fig:delta_DT_mass_frac} which shows that such events typically induce a significant morphological change.  Note that the mass ratio is defined in terms of the total mass of the satellite to the stellar mass of the host at the redshift of the merger (constrained to be $z < 2$).

It is evident from Fig.~\ref{fig:DT_mass_frac} that there is no strong correlation between D/T at z=0 and the mass ratio of the last massive merger.  The Spearman rank correlation coefficient is weak, with $r=-0.13$.  Note that this result is insensitive to our choice of definition of the mass ratio.  For example, if we define the mass ratio in terms of the stellar mass of the host at $z=0$ instead of at the redshift of the merger, we find a similarly poor correlation with the present-day morphology.

Taken together, Figs.~\ref{fig:delta_DT_mass_frac} and \ref{fig:DT_mass_frac} therefore strongly suggest that the morphology is altered (again) following the last massive merger (i.e., through disc re-growth).

Note that some present-day disc galaxies may have had multiple massive mergers (i.e with ${\rm M}_{\rm sat}/{\rm M}_{\rm star, host} >1$) during their lifetime. Some of these mergers could have higher mass ratios than the last massive mergers. However, given the poor correlation between the present day morphologies and the properties of the last massive merger, it is unlikely that the conclusions of Fig. \ref{fig:delta_DT_mass_frac} can change regarding correlations with the properties of earlier mergers (on the contrary, in that case we will expect an even weaker correlation).

\begin{figure}[t]
\begin{center}
\includegraphics[width=0.6 \columnwidth]{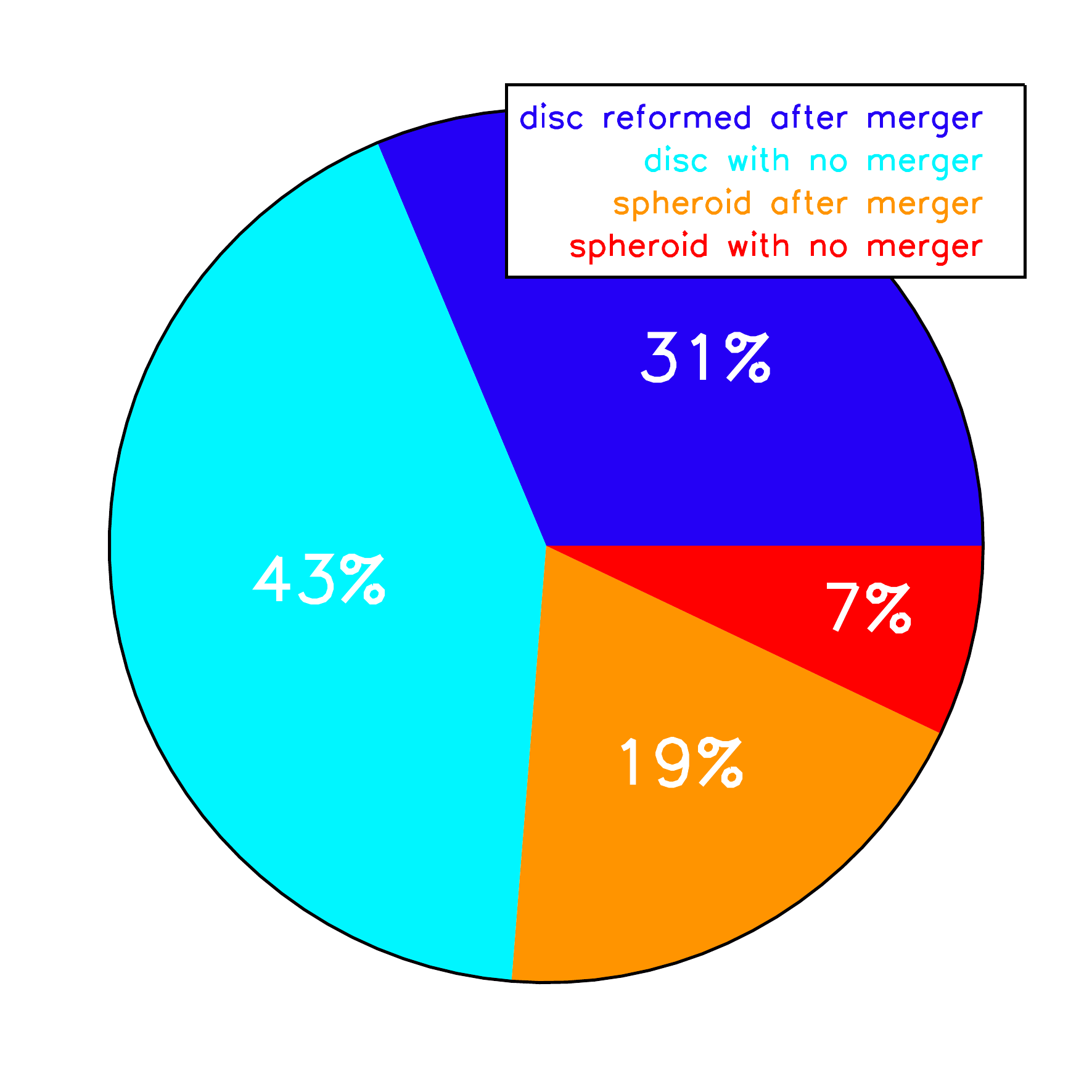} 
\includegraphics[width=0.6 \columnwidth]{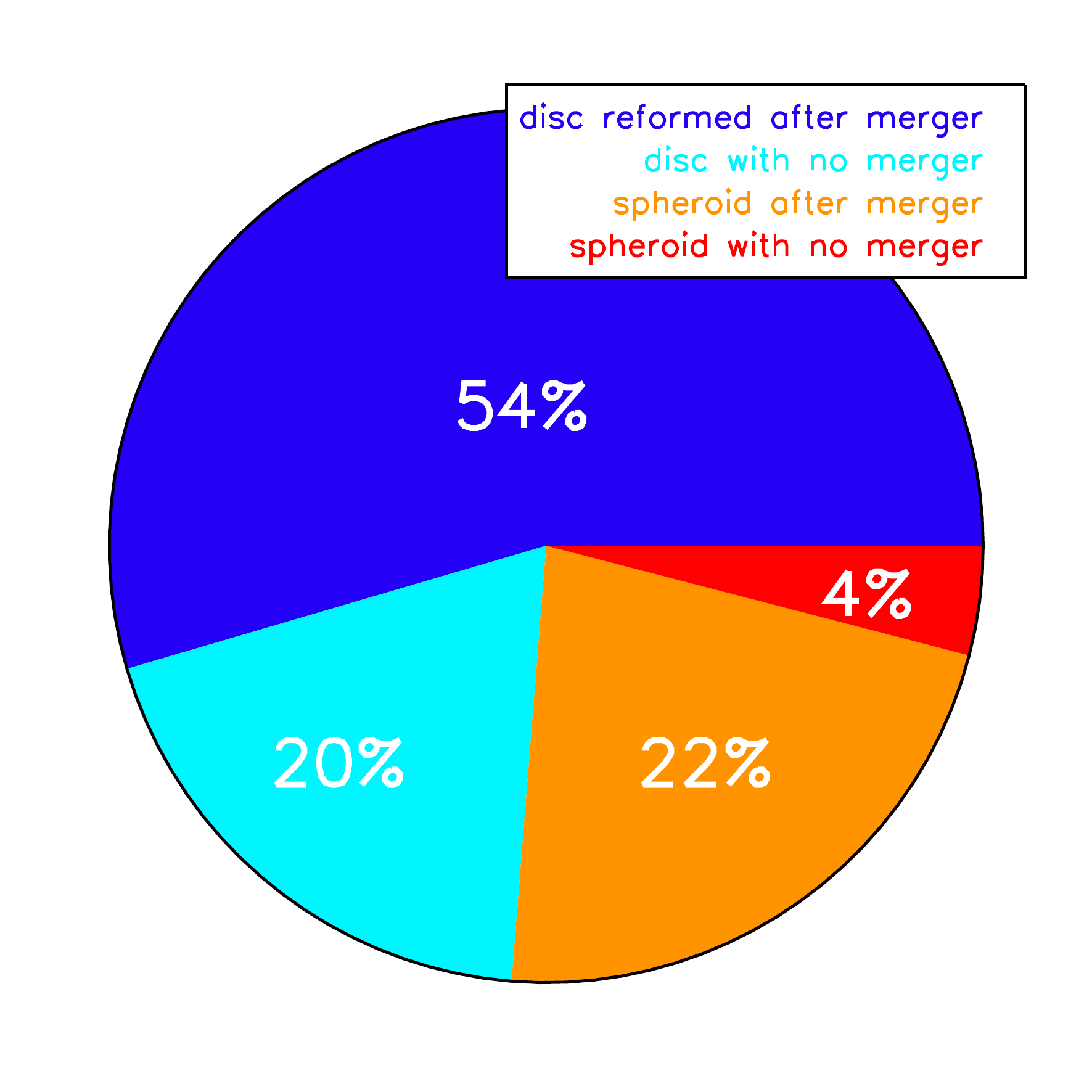}
\caption{The distribution of galaxies, both with and without massive (${\rm M}_{\rm sat,tot} > {\rm M}_{\rm star,host}$) mergers since $z_{\rm merge} \leq 1$ (top) and since $z_{\rm merge} \leq 2$ (bottom), respectively. The four categories are: galaxies that are discs today and had a merger (blue); galaxies that are discs and did not undergo a merger (cyan); galaxies that are spheroids after a merger (orange); and spheroids that did not have a merger (red).}
\label{fig:pie_chart}
\end{center}
\end{figure}

What is the prevalence of the disc re-growth after mergers? In Fig. \ref{fig:DT_mass_frac} we looked only at the fate of galaxies that underwent massive (${\rm M}_{\rm sat,tot} > {\rm M}_{\rm star,host}$) mergers - albeit, given the large redshift range under consideration, this comprises the majority of our galaxies. In Fig.~\ref{fig:pie_chart} we capture the morphological changes of all galaxies since a given time. As before, we divide galaxies in two categories, those that had a massive merger (${\rm M}_{\rm sat,tot} > {\rm M}_{\rm star,host}$) and those that have not. The top panel shows the morphological changes of galaxies since $z_{\rm merge} \leq 1$ (or lookback time of $\simeq 7$ Gyr). With these criteria, $31\%$ of all galaxies today are discs reformed after massive mergers. In contrast, only $19\%$ of galaxies became (or remained) spheroids after a massive merger. A significant fraction, $43\%$, are present-day discs that had quiescent histories, i.e no massive mergers since $z=1$, while $7\%$ are spheroids that also had no recent massive mergers. The bottom panel of Fig.~\ref{fig:pie_chart} shows that if we set the massive merger time scale further back to $z=2$ (lookback time of $\simeq 10$ Gyr), the disc re-growth process is even more prevalent: $54\%$ of all galaxies today are discs that had a massive merger. The rest, $22\%$ are spheroids today which had a massive merger, $20\%$ are discs that had no merger and 4\% are spheroids that had no merger. These plots show clearly the resilience of disc galaxies in face of disruptive mergers.

\subsection{Dependence on cold gas fractions}
\label{sec:gas_frac_DT} 

As discussed in the Introduction, it has been proposed that the fraction of cold gas available at the time of mergers, can predict the (re)formation of disc galaxies after massive mergers.  Here we revisit the relation between cold gas fractions and galaxy morphology (D/T) by tracking both self-consistently in a cosmological context.  Note that here we define the cold gas mass fraction, $f_{\rm gas}$, as the ratio of gas mass to total mass\footnote{We have also tried defining the gas mass fraction just in terms of the baryonic component (i.e., $f_{\rm gas}={\rm M}_{\rm gas}/{\rm M}_{\rm (gas+stars)}$), as in some previous studies (e.g., \citealt{hopkins09}), and find qualitatively similar results.} within the central 20 kpc.

\begin{figure}
\includegraphics[width=0.995\columnwidth]{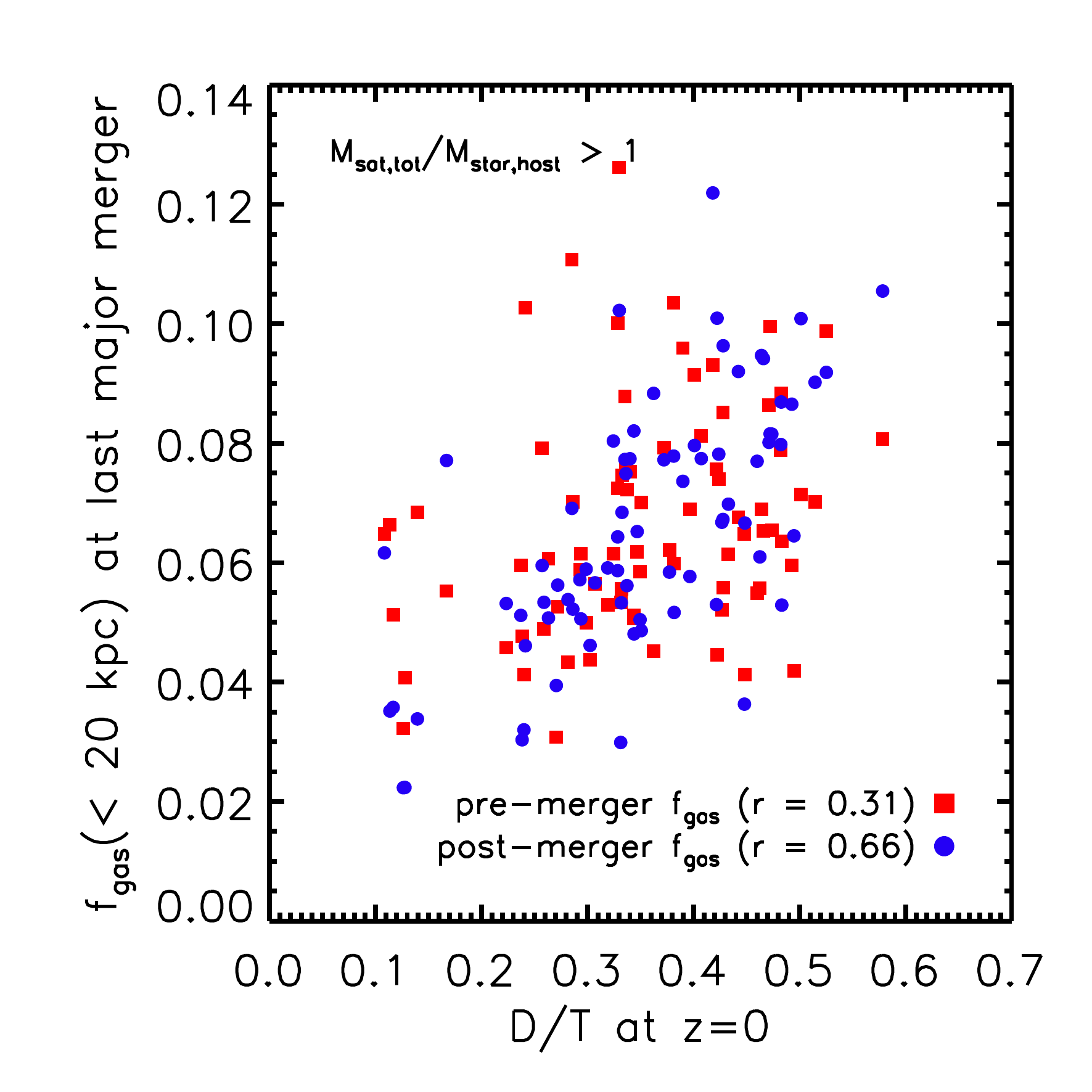}
\caption{The correlation between galaxy morphologies (D/T)  at $z=0$ and $f_{\rm gas}(<20$~kpc) at the time of the most massive $>{\rm M}_{\rm star, host}$ merger. The gas fractions after the merger (filled blue circles) exhibit a stronger correlation with D/T (z=0) than the fractions computed prior to the merger (filled red squares). The Spearman rank coefficient for the post-merger $f_{\rm gas} -$ D/T (z=0) correlation is 0.66, while that of the pre-merger $f_{\rm gas}$ $-$ D/T (z=0) correlation is 0.31.}
\label{fig:fgas_DT}
\end{figure}

We examine the relation between present-day `diskiness' and gas fractions on a per-galaxy basis. Fig.~\ref{fig:fgas_DT} shows the correlation between the morphologies of present-day galaxies and the fraction of gas $f_{\rm gas}$ present within the central 20 kpc of the host at the time of the most massive merger for major merger cases (i.e., mass of the satellite exceeded ${\rm M}_{\rm star, host}$ at collision).  We experiment with using the gas fraction defined both immediately pre- and post-merger.  In both cases, there are significant correlations with present-day morphology.  The correlation is particularly strong (with a Spearman rank correlation coefficient of 0.66) when using the gas mass fraction measured immediately post-merger.

We note that using the same \gimic simulations, \citet{sales12} found that galaxy morphologies at $z=0$ depend on the fraction of the {\it hot} gas in the host galaxy rather than that of the cold gas. However, this is likely the result of the inclusion of much more massive galaxies in their sample (${\rm M}_{200} \simeq 1-3 \times 10^{12} {\rm M}_{\odot}$).  More massive galaxies naturally have higher $f_{\rm hot}$, but in \gimic, because they are likely to be over-cooled, they can also have artificially compact/robust discs with high D/T ratios.

Where does the gas that fuels disc (re)growth come from?  Most of the gas participating in star formation has been accreted smoothly, rather than being brought in by satellites. For example, in a sub-set of the OWLS simulations similar to \gimic \footnote{Their model \textsc{REF-L050N512} uses the same prescriptions for feedback \citep{dallavecchia08} and the same code. The mass-loading factor is a factor of two higher in \gimic, however this does not change the star formation histories of galaxies significantly (see \citealt{sales10}).}, \citet{vandevoort11} find that the gas associated with substructure comprises only $\sim 10\%$ to the total gas budget in $\sim 10^{12} {\rm M}_{\odot}$ mass galaxies (see their Fig.~3). This is likely the result of the strong stellar feedback implemented in both of these simulations which unbinds a large fraction of gas from lower mass dark matter subhaloes. Other studies obtain similar results. In particular, \citet{brooks09} find that in sub-$L_{*}$ galaxies, cold flow gas accretion is responsible for star formation in the disc throughout the galaxy's lifetime. These results suggest a direct relation between the fraction of cold gas and disc morphologies.

Understanding the mechanisms by which cold gas operates on the disc structure is also important. In principle, there are two distinct channels: one in which its acts mainly toward (re)forming the stellar discs after the mergers (which is what we have mainly focused on so far), and the other in which the cold gas mitigates the tidal impact and increases the resilience of discs to mergers.  An example of direct mitigation is provided by \citet{moster10} who show that the cold gas in the disc can absorb part of the tidal energy of the merger into its thermal energy (and radiate it later) and consequently, the stellar disc maintains a post-merger scale-height consistent with the observations. However, this mechanism is unlikely to be efficient in the case of massive mergers (indeed, we have already shown that 1:1 satellite-to-galaxy stellar mass mergers usually induce very large changes to the kinematic properties of the central galaxy), and possibly also in the case of many minor mergers following in short succession.  

Another suggestion is that the prevalence of present-day discs is achieved via efficient feedback which efficiently removes gas from systems until after the main merger activity has subsided ($z\sim 1$), the gas is then re-accreted and forms a disc \citep{weil98}.  While efficient feedback to prevent excessive star formation at high redshift is certainly important, it cannot by itself be the whole story.  For example, it cannot easily explain the observations of large numbers of disc galaxies out to $z\sim 2$ (see \citealt{vanderkruit11} and references therein) or the fact that a fraction of the Milky Way thin disc has formed before $z\sim1$ \citep{wyse01,haywood13}.  

If mitigating mechanisms are indeed generally inefficient, the explanation for the present-day abundance of disc galaxies must rely on the efficient star-formation activity which promotes the re-growth of stellar discs \citep{hammer07}. This process, which entails successive morphological transformations during the lifetime of a galaxy, lies at the basis of forming large discs in \gimic and it will be investigated in more detail in Section \ref{sec:pathways}.

\subsection{Dependence on time since the last massive merger}
\label{sec:merger_hist}
 
\begin{figure}
\includegraphics[width=0.995\columnwidth]{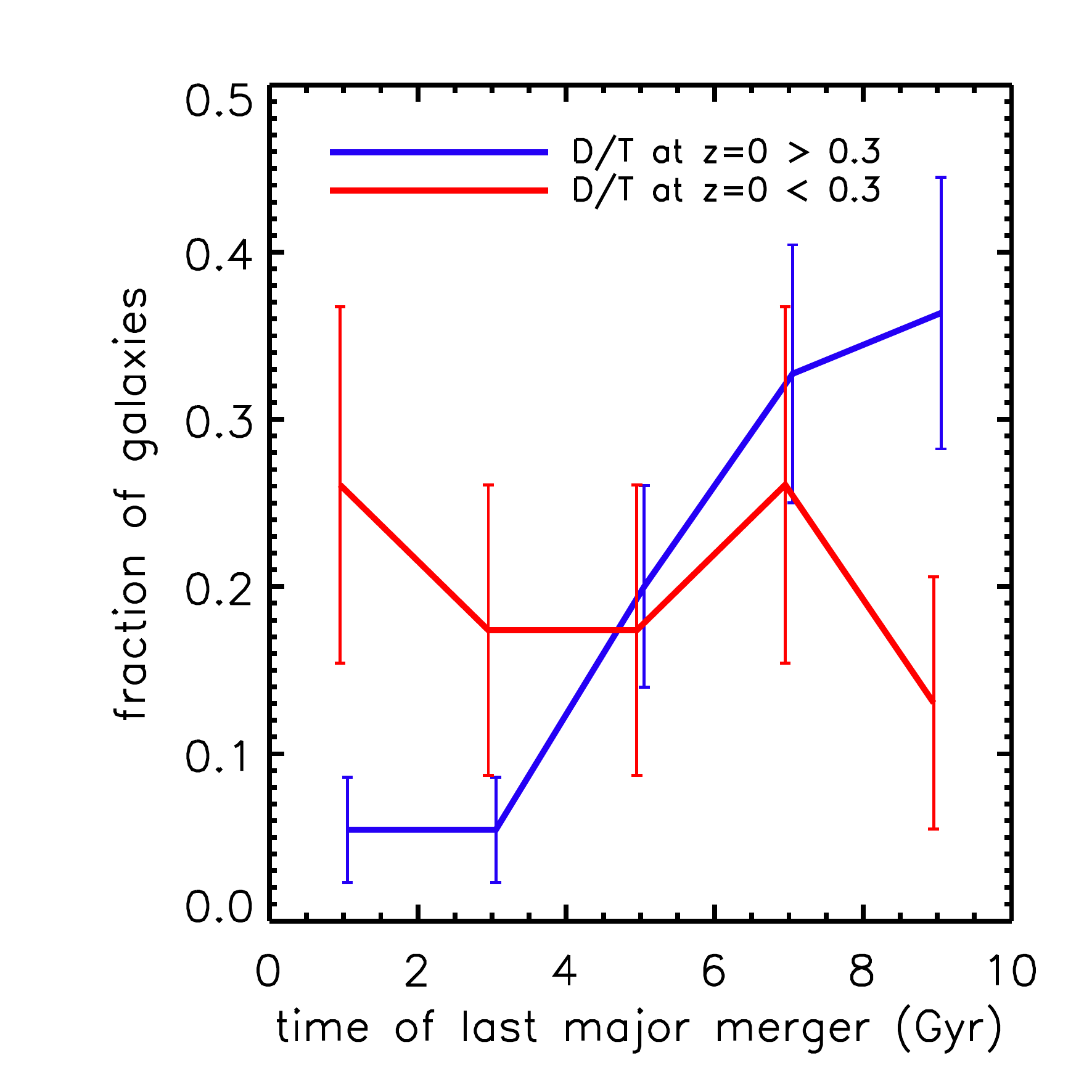}
\caption{The distribution of the times since the last massive (${\rm M}_{\rm sat}/{\rm M}_{\rm star, \rm  tot} > 1:1$) merger for disc (D/T $\ge 0.3$; blue curve) and spheroid (D/T $\le 0.3$; red curve) galaxies at $z=0$. The error bars represent Poisson uncertainties.  M$_{\rm sat}$ is measured when the satellites first cross $r_{\rm merge}=20$~kpc (see text).  Present-day disc galaxies have had more quiescent recent histories compared to spheroids of the same stellar mass.}
\label{fig:tmm}
\end{figure}

Another key factor which can influence the (re)formation of discs is the amount of time elapsed between the last massive merger and the time of observation (the present day in this case).  Fig.~\ref{fig:tmm} shows the distribution of the time of the last massive mergers ($> {\rm M}_{\rm star, host}$) for present-day galaxies categorised broadly into disc and spheroid galaxies.  Most disc galaxies have had their last massive merger prior to $z\sim1$ ($\simeq 7-8$~Gyr ago). Galaxies with recent mergers tend to become spheroids, especially if $f_{\rm gas}$ at the time of merger is low (which is likely at $z<1$).  Note that this figure does not differentiate between the mass ratios of different mergers.  Of course, galaxies which experience mergers of higher mass ratio are more likely to become spheroids (see Fig.~\ref{fig:delta_DT_mass_frac} and Section \ref{sec:pathways}). These results suggest that there are important differences in the merger histories of disc and spheroid galaxies of fixed (stellar) mass.

\begin{figure*}
\centering
\includegraphics[width=1.95\columnwidth]{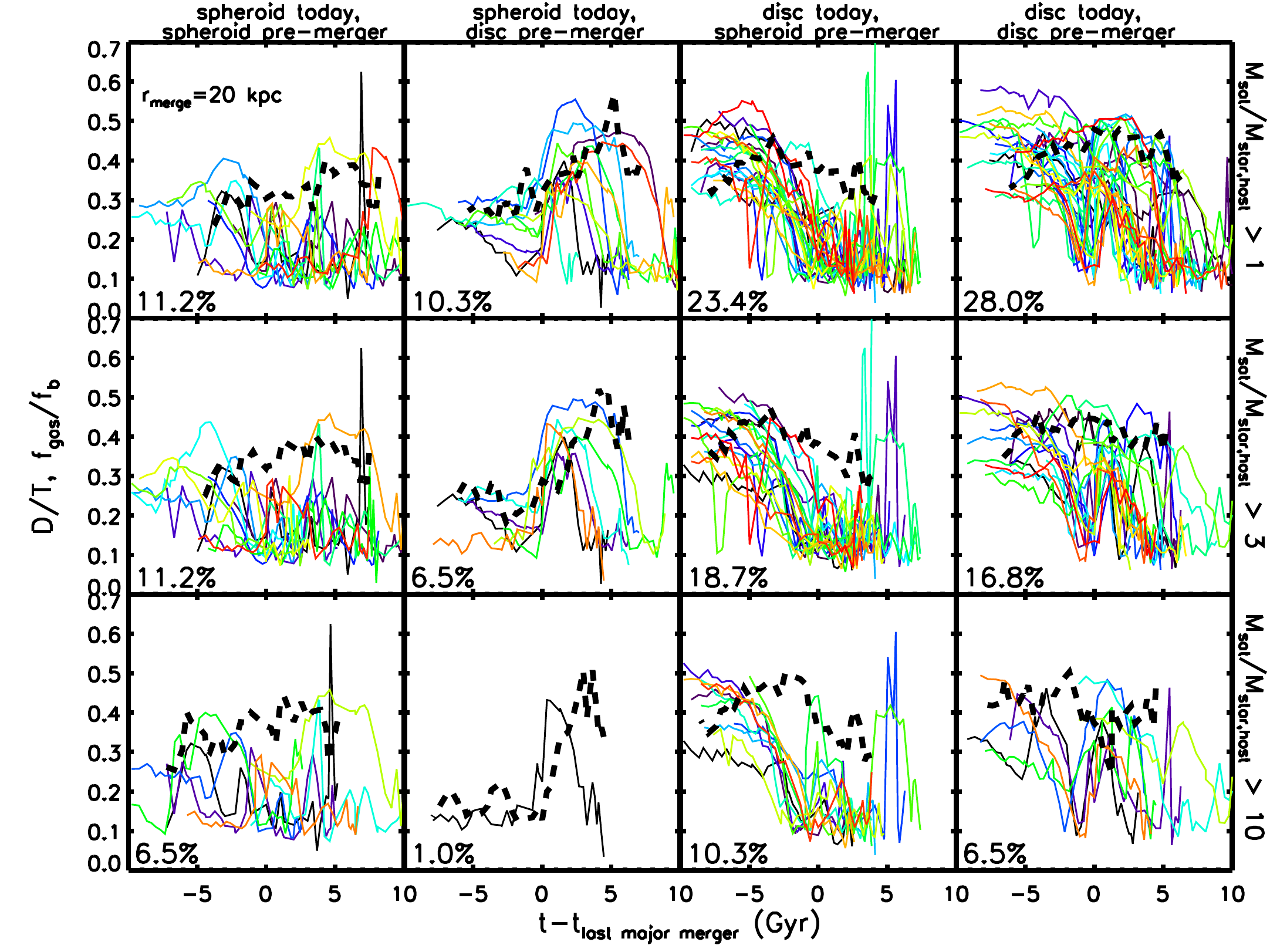}
\caption{Various D/T trajectories ("pathways") for forming disc galaxies and spheroids including all possible morphological transformations (the four columns): spheroid today--spheroid pre-merger, spheroid today--disc pre-merger, disc today--spheroid pre-merger, and disc today--disc pre-merger.  The percentage at the bottom left of each panel indicates the fraction of galaxies in the original sample which undergo these transformations.  The three rows from top to bottom show galaxies with last massive mergers above the thresholds: $>{\rm M}_{\rm star, \rm host}$, $>3 {\rm M}_{\rm star, \rm host}$ and $>10 {\rm M}_{\rm star, \rm host}$.  The coloured lines in each panel show the D/T trajectories of individual galaxies and the dashed black lines in each panel indicate the median $f_{\rm gas}/f_b$ within $20$~kpc.  $t=0$ represents the time of the last massive merger, with the mass ratio threshold indicated on the right vertical axis. Approximately half of all galaxies today (in this mass range) are both disc-dominated and have had a massive merger at some point in their past.  Approximately half of these galaxies were discs prior to the major merger and re-grew their disc afterwards (while the other half were spheroids prior the last massive merger and grew discs later on).
}
\label{fig:DT_history}
\end{figure*}

Consistent with Fig.~\ref{fig:DT_mass_frac}, Fig.~\ref{fig:tmm} also indicates that quiescent merger histories are not a prerequisite for the formation of large disc galaxies.  Recall that all disc galaxies in this plot have non-quiescent merger histories, i.e. mergers $> {\rm M}_{\rm star, host}$.   Secondly, a non-negligible fraction of these have had their last $> {\rm M}_{\rm star, host}$ since $z \sim 1$. The likelihood of disc galaxies forming via more quiescent merger histories than shown in Fig.~\ref{fig:tmm}, e.g. with no mergers as massive as ${\rm M}_{\rm star, host}$ since $z=2$, is low, $<20\%$ (see the dashed orange line in Fig.~\ref{fig:merger_frac}).  Therefore, the formation of the Milky Way via a very quiescent merger history cannot be ruled out in a $\Lambda$CDM model, even though it is statistically unlikely\footnote{It is interesting to note here that a future merger with the Large Magellanic Cloud will not be far from a 1:1 case when comparing the total mass of the LMC to the Milky Way's stellar mass.} (the likelihood is even lower, $<10\%$, if we require that Milky Way experiences only mergers $\leq 0.5 {\rm M}_{\rm star, host}$, i.e.$\leq {\rm M}_{\rm star, disc}$ since $z=2$; see the red dashed line in Fig.~\ref{fig:merger_frac}).  On the other hand, quiescent merger histories, by themselves, do not guarantee the emergence of galaxy discs.  The misalignment of angular momenta of the infalling gas may transform a disc galaxy into a spheroid even in the absence of mergers \citep{sales12}. 

\subsection{Cold gas fractions and merger time combined: The diversity of pathways for forming disc galaxies.}
\label{sec:pathways}

So far we have studied the role of cold gas fractions and merger histories separately and have found both to be important.  In reality these two factors are inter-related. Fig.~\ref{fig:DT_history} encapsulates the various likelihoods of forming disc ($\textrm{D/T} >0.3$) and spheroidal ($\textrm{D/T}<0.3$) galaxies, as a function of the mass ratio of their last massive ($>{\rm M}_{\rm star, \rm host}$) mergers (the three rows) and as a function of the average $f_{\rm gas}(<20$~kpc) (dashed black lines in each panel).  As before, we focus only on galaxies with non-quiescent merger histories, which are typical in a hierarchical cosmology.  The mass thresholds for these mergers are, from top row to bottom row, $>{\rm M}_{\rm star, \rm host}$, $>3 {\rm M}_{\rm star, \rm host}$ and $>10 {\rm M}_{\rm star, \rm host}$. The coloured lines in each panel show the D/T histories of individual galaxies, with the time being measured with respect to the time of the last massive merger for that galaxy, where only mergers with mass ratios above the threshold corresponding to each row are considered. (Note that, in this tally, some mergers can be captured in more than one panel, e.g. a merger with $>3 {\rm M}_{\rm star, \rm host}$ may appear  in a panel with $>{\rm M}_{\rm star, \rm host}$, if that was the last massive merger experienced by the galaxy).

The four columns from left to right tally all possible morphological transformations between the time just prior to the last massive merger and $z=0$:  spheroid - spheroid, disc - spheroid, spheroid - disc and disc - disc. The percentage at the bottom left of each panel represents the fraction of galaxies relative to the total number of systems in the original sample which undergo these transformations (note that the total number also includes galaxies with quiescent histories or those with non-massive mergers). For example, the two rightmost panels on the top row show that about $50\%$ of all galaxies experience a merger $>{\rm M}_{\rm star, \rm host}$ since $z=2$ and yet are disc-like today. About $23\%$ of sample galaxies were spheroids before the last $>{\rm M}_{\rm star, \rm host}$ merger and are disc-like today, while a similar fraction ($\sim 28\%$) were disc-like before their last $>{\rm M}_{\rm star, \rm host}$ merger and are still disc-like today (however, these are usually not the same discs). The formation of discs after the mergers is more likely to occur when the fraction of cold gas is high, $f_{\rm gas}/f_b >0.4$ (where $f_b = \Omega_b/\Omega_m$), irrespective of whether galaxies were spheroids or discs before the merger. 

Disc galaxies sometimes manage to form even in the case of very massive mergers, albeit the likelihood of this happening is very low. For example, about $50\%$ of sampled galaxies have mergers ${\rm M}_{\rm sat} > 3 {\rm M}_{\rm star, host}$ and most of them become disc galaxies by $z=0$. About $30\%$ of all galaxies in our mass range may form discs this way (the two rightmost panels in middle row).  Interestingly, disc galaxies can form even after mergers $> 10 {\rm M}_{\rm star, host}$ (about $16.8\%$ of all sample galaxies). Again, the determining factor is the high gas fraction ($f_{\rm gas}>0.4$) at the time of merger and thereafter. 

In summary, quiescent merger histories are certainly not a prerequisite for the formation of disc galaxies. Most galaxies undergo mergers massive enough to significantly affect/destroy stellar discs (${\rm M}_{\rm sat} \geq {\rm M}_{\rm star, host}$), however new discs can often form afterwards, even when these mergers are relatively recent (e.g. $z \sim 1$, recall Fig.~\ref{fig:tmm}). These results run counter to the expectations from collisionless studies in which mergers this massive will permanently turn discs into spheroids. 

We note that the diversity of merger histories of disc galaxies has been noted before by \citet{martig12}, who found little correlation between the morphologies of disc galaxies today and of their progenitors at $z\sim1$. These authors also note that the thinnest disc galaxies in their sample are formed in quiescent histories and suggest a quiescent formation scenario for the Milky Way.

\section{The disc re-growth scenario in the context of observations}
\label{sec:imprints}

As we have argued so far, disc galaxies can form in the context of violent mergers. This does not preclude the disruption of discs by the incoming mergers (quite the contrary, we have shown clear morphological changes result from these mergers). So a process of disc re-growth has to happen, facilitated by the abundance of cold gas around these galaxies. Here we examine some of the observational evidence that may support this scenario. 

\subsection{The mass growth of discs and the discs stellar ages}

\begin{figure}
\includegraphics[width=0.995\columnwidth]{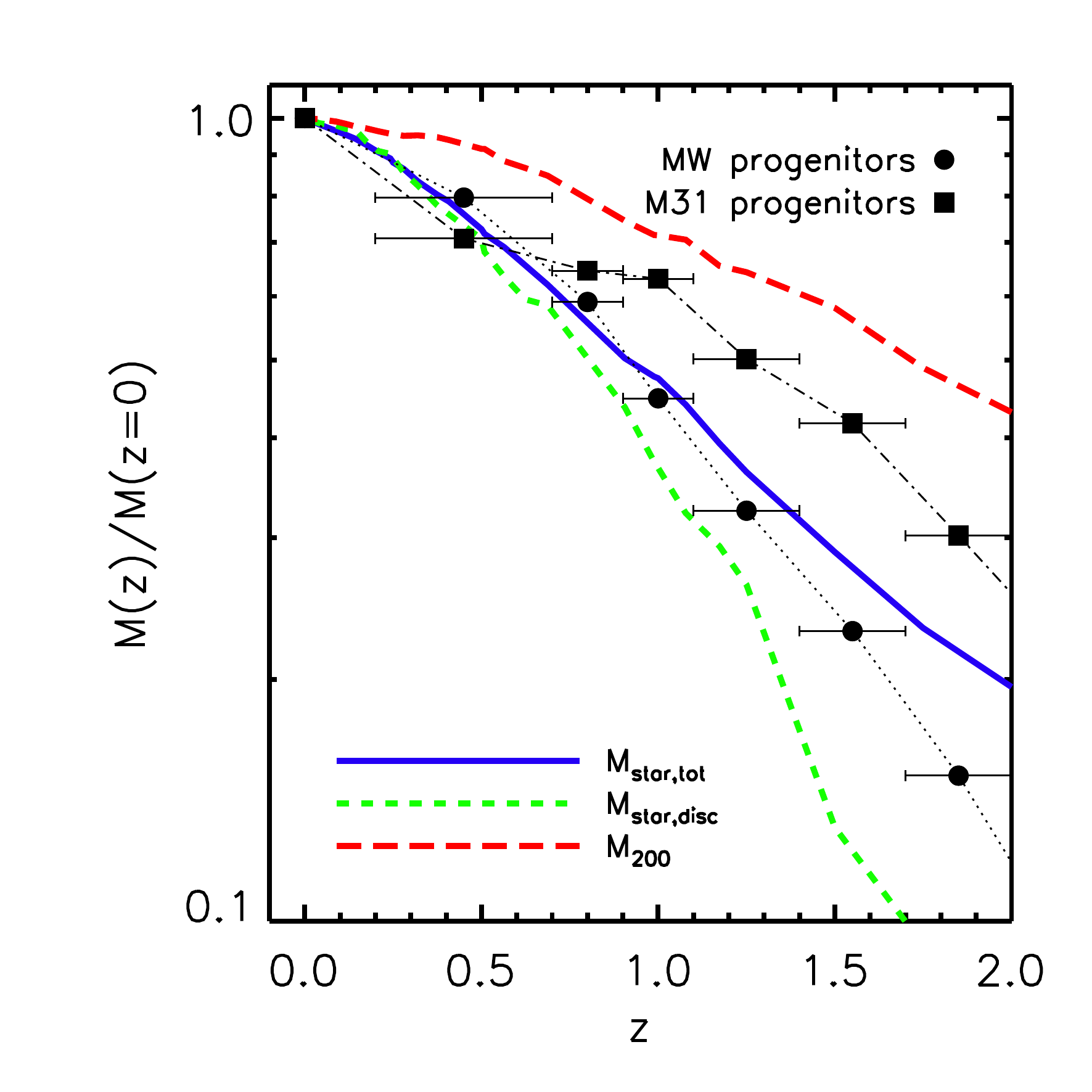}
\caption{The growth in stellar mass in the simulated disc galaxies, compared to that inferred for Milky Way- and M31-like progenitors from the ZFOURGE/CANDELS survey \citep{papovich15}.  Both the simulated and observed galaxies effectively double their stellar mass content since $z\sim1$ (i.e., post the main merger period).  The growth of stellar discs (dashed green curve) tracks the total stellar mass growth since $z\sim1$ in the simulations.
}
\label{fig:disc_growth}
\end{figure}

First, there is evidence that stellar discs grow significantly in mass at recent times.  Studies that examined the stellar mass growth of typical disc galaxies (by looking at their likely progenitors) have found that they typically double their stellar mass since $z\sim1$ \citep{bell05,hammer05,patel13,vanDokkum13,papovich15}.  Thus, there is clearly significant (re)growth post the main merger period, in support of the general ideas advocated above.  In Fig.~\ref{fig:disc_growth} we make a more quantitative comparison -- the stellar mass growth of the simulated disc galaxies with that recently inferred for Milky Way- and M31-like progenitors from ZFOURGE/CANDELS data by \citet{papovich15}.  These authors traced the evolution of systems with present-day stellar masses of $5\times10^{10} M_\odot$ (MW-like) and $10^{11} M_\odot$ (M31-like) back in redshift, effectively identifying analogues of their progenitors in the ZFOURGE/CANDELS survey as systems with the same comoving number density as the $z=0$ systems.  Overall the growth of stellar mass\footnote{Note that this is the instantaneous stellar mass at a given redshift $z$; i.e., it is {\it not} equivalent to the integral of the star formation history from high redshift down to $z$, as some stellar mass is lost over time due to stellar evolution (both in the simulations and in nature).} in the simulated galaxies is reasonably compatible with the observations.  Furthermore, we note that the growth of the mass of the stellar discs in the simulations tracks the total stellar mass growth closely below $z\sim1$, indicating that this is the dominant period of disc formation in the simulations.

\begin{figure}
\includegraphics[width=0.995\columnwidth]{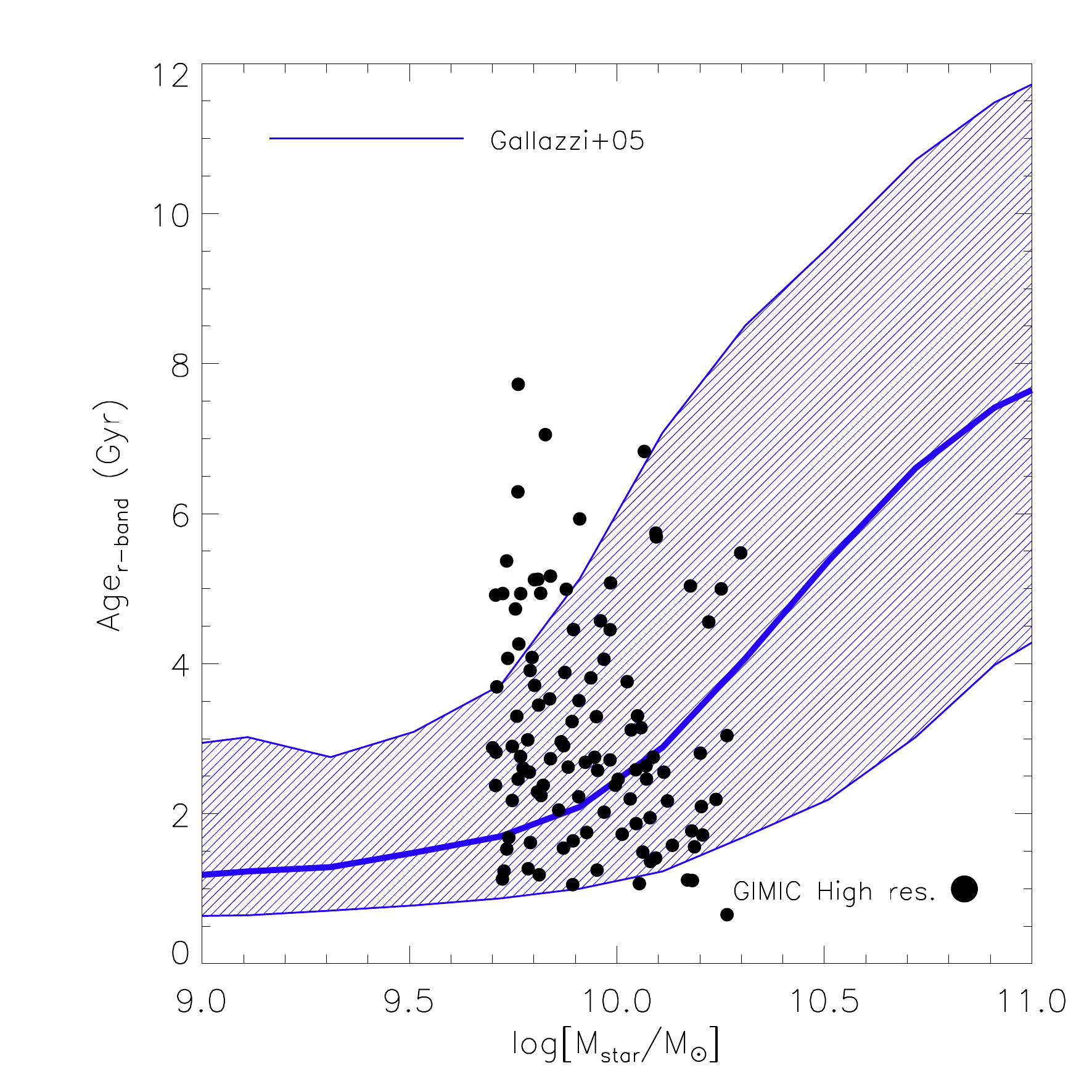}
\caption{The the ($r$-band) luminosity-weighted ages of the simulated discs (black circles) compared with the data of \citet{gallazzi05} (shaded region).  As typical disc galaxies exhibit merger activity until $z\sim1$, the stellar discs are generally young/intermediate-age, although recent/ongoing star formation biases the luminosity-weighted ages towards lower values (see text).}
\label{fig:ages_disc}
\end{figure}

The median age of the disc stars can be another indication. Since the merger activity of most disc galaxies continues until $z\sim1$, a typical stellar disc that re-forms post merger will contain a significant fraction of young and/or intermediate age stars. The median ages of our simulated discs are about $5-6$~Gyr. This compares well with the inferred half-age of the Milky Way disc, of  $\sim 6.9$~Gyr \citep{aumer09}. The luminosity-weighted disc ages are, as expected, younger for systems with ongoing star formation. Fig.~\ref{fig:ages_disc} shows a comparison between the $r$-band luminosity-weighted ages in our simulated discs and the data of \citet{gallazzi05}. Similar younger ages are also obtained in a local sample of disc galaxies by \citet{yoachim08}. 

Of course, not all disc galaxies in the observational sample may have underwent massive mergers. However, the simulated sample shown in this plot includes also {\it all} disc galaxies, some of which also did not undergo massive mergers. The premise of this comparison is that within the context of the $\Lambda$CDM model, the majority of disc galaxies observed at $z=0$ experience such mergers, as shown by our simulations.

\section{Conclusions}
\label{sec:concl}


Reconciling the ubiquity of disc galaxies at the present day with the prevalence of (apparently) disc-destroying mergers that galaxies face in the $\Lambda$CDM cosmological context has been a subject of intense research for several decades.  While progress was undoubtedly stalled due to issues linked to overcooling/inefficient feedback (leading to overly spheroidal galaxies) in earlier simulations, the current generation of cosmological hydrodynamical simulations can now, generally speaking, successfully produce large populations of disc galaxies whose properties compare favourably to observations (e.g., \citealt{vogelsberger14,hopkins14,schaye15}).  The implication of this development is that the $\Lambda$CDM cosmological model is a perfectly viable setting for the emergence of a large disc galaxy populations.  The aim of the present study is to help elucidate the physics that enables galaxies to retain (or re-form) their discs, and to clarify the somewhat unclear picture that has emerged from studies based on previous zoomed simulations of small numbers of systems.  

The main results our study can be summarised as follows:

\begin{itemize}

\item A quiescent merger history is not a prerequisite for the existence of disc galaxies today. We find that more than half of simulated galaxies had at least one merger with a mass $\geq {\rm M}_{\rm star, \rm host}$ sometime since $z=2$ and yet are disc galaxies today, while a third and one sixth of galaxies experienced mergers $\geq 3 {\rm M}_{\rm star, \rm host}$ and $\geq 10 {\rm M}_{\rm star, \rm host}$, respectively, and still they have disc morphologies today (see Fig.~\ref{fig:DT_history}).  One half of the M$_{\rm sat}>{\rm M}_{\rm star, host}$ impacts occur at look-back time of $\le 6-7$~Gyr. The pathways by which disc galaxies can emerge from merger-induced morphological transformations are remarkably diverse.  Most galaxies undergo relatively frequent morphological transformations by the present-day.

\item The majority (approximately 80\%) of galaxies in the mass range under consideration undergo massive mergers (with ${\rm M}_{\rm sat} \geq {\rm M}_{\rm star, host}$) since $z=2$.  Such mergers typically induce large changes in the morphology of disc-dominated galaxies. In spite of this, there is essentially no correlation between the present-day morphology of galaxies and the mass ratio of the last massive merger, indicating that the morphology is altered again (i.e., the disc re-grows) post-merger. Disc re-growth occurs in more than half of all galaxies over the past $\sim 10$~Gyr and for about a third of galaxies over the past $\sim 7$~Gyr.

\item Galaxies with high gas fractions either immediately pre- or (especially) post-merger can reform their discs, particularly if the last massive merger occurred at $z\sim1$ or earlier.


\end{itemize}

The results presented here apply to `normal' systems with present-day masses of $\sim10^{10} M_\odot$, which is the roughly the mass scale that has the largest fraction of systems with significant disc components in the local Universe and where the \gimic simulations reproduce the observable properties of galaxies relatively well.  These simulations lack the resolution to explore the morphologies of galaxy's significantly below this mass scale, and they lack high-efficiency feedback (i.e., from AGN) required to prevent overcooling in significantly more massive systems.  Note, however, for dwarf galaxies it is already known that quiescent formation is not necessary to obtain stellar discs (see, for example \citealt{brook11}). It is clearly of interest to perform the kind of analysis done above the Milky Way mass scale, in an attempt to understand why disc formation is so inefficient on those scales. The advent of a new generation of simulations, such as Illustris, EAGLE and Horizon-AGN (which extend to larger volumes and include AGN feedback), represent promising tools for pursuing this line of research for more massive systems, while much higher-resolution `zoom' simulations (e.g., FIRE, APOSTLE) are best suited to investigate the morphologies of lower mass systems. 

\begin{acknowledgements}
The authors thank the referee, Fabio Governato, for his constructive comments and suggestions that helped improved the paper. They also thank Peter Yoachim for providing the observational data in Fig.~1 in electronic format, Peter Behroozi for helpful comments and the EAGLE team for providing a comparison sample from their simulations.
During part of this project ASF was supported by a Royal Society Dorothy Hodgkin Fellowship. IGM is supported by a STFC Advanced Fellowship. RAC is a Royal Society University Research Fellow. AMCLB acknowledges support from an internally funded PhD studentship at the Astrophysics Research Institute of Liverpool John Moores University. The authors thank their collaborators within the Virgo Consortium for support with the development of the original \gimic simulations. The two high resolution runs discussed here were performed on the Darwin Supercomputer at the University of Cambridge. 
\end{acknowledgements}

\bibliographystyle{pasa-mnras}
\bibliography{mybib}

\begin{appendix}

\begin{figure*}
\centering
\includegraphics[width=0.99\columnwidth]{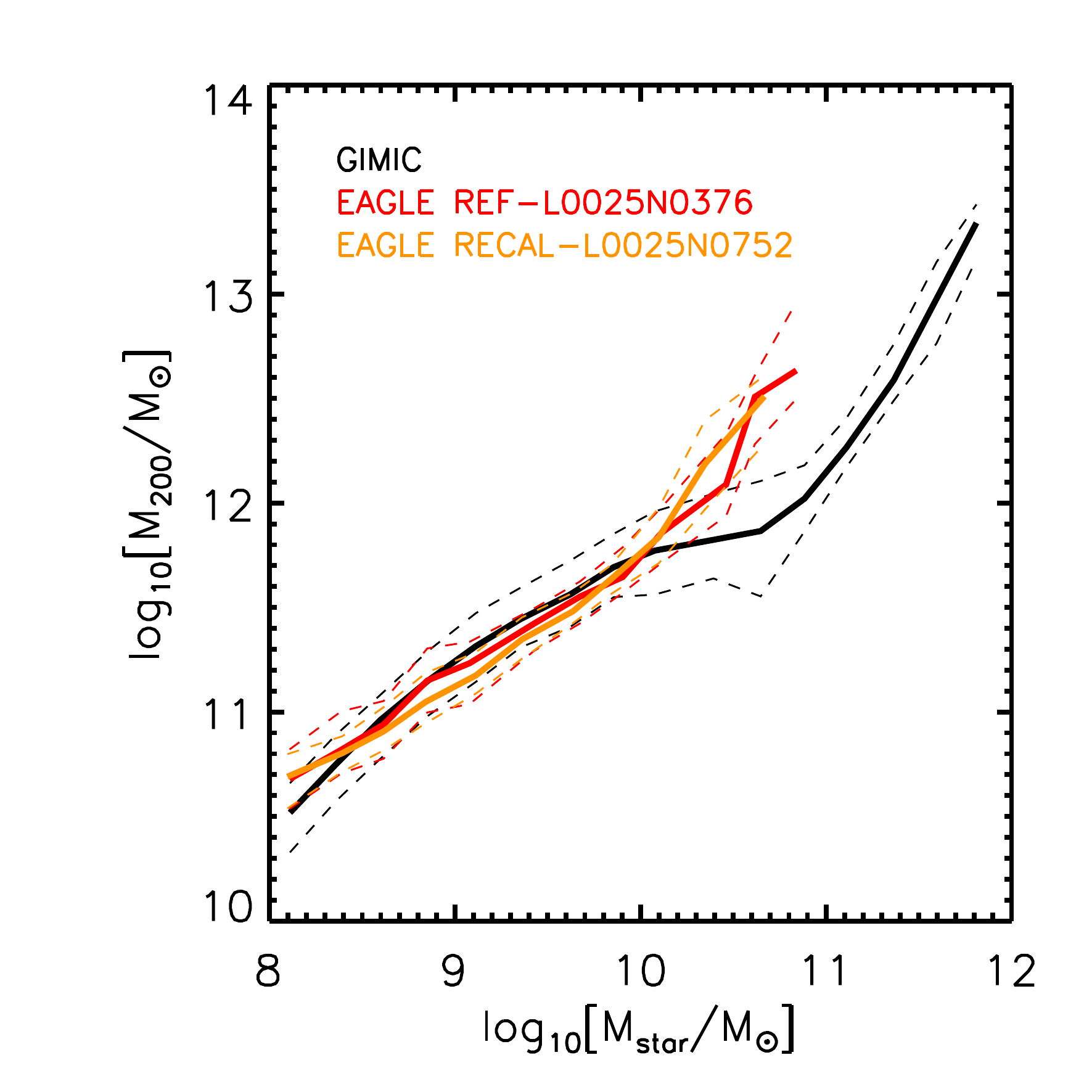}
\includegraphics[width=0.99\columnwidth]{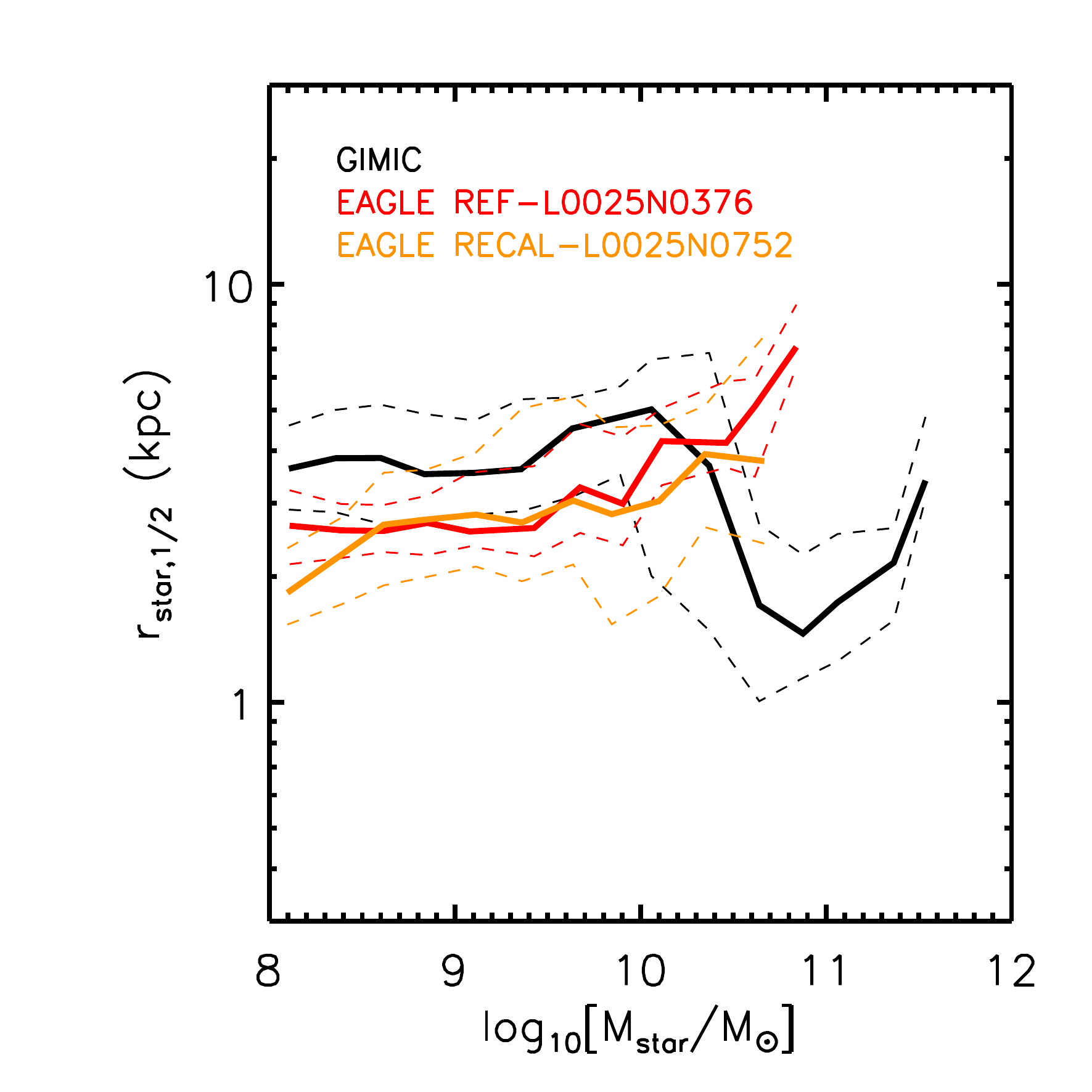}
\caption{Comparison of the present-day stellar mass$-$halo mass relations ({\it left}) and stellar mass$-$half-mass radius relations ({\it right}) of \gimic\ with the recent EAGLE simulations.  The half-mass radius, $r_{\rm star,1/2}$, is defined as the radius which encloses half of the total stellar mass bound to the galaxy's halo (excluding the stellar mass of any satellites).  At stellar masses of less than approximately $10^{10}$ M$_\odot$, the simulations predict similar stellar mass$-$halo mass relations, while the \gimic\ galaxies have slightly larger half-mass radii.  At higher stellar masses, the \gimic\ simulations suffer from overcooling, resulting in higher stellar masses for a given halo mass and galaxies that are too compact (see also \citealt{mccarthy12b}).}
\label{fig:Mstar_GIMIC_EAGLE}
\end{figure*}

\section{Comparison to EAGLE}
\label{sec:appendix_eagle}

Here we compare the $z=0$ stellar mass$-$halo mass relations, stellar mass$-$size relations, and the morphologies of the \gimic\ galaxies with those derived from the recent EAGLE simulations \citep{schaye15}.  For the EAGLE simulations, we use the fiducial reference (REF) model at the fiducial resolution (which is similar to the high-res.~\gimic\ runs that we use here) and the recalibrated (RECAL) model for the higher-resolution EAGLE runs (which are a factor of 8/2 better mass/spatial resolution compared to the fiducial resolution runs, respectively).  We select EAGLE galaxies in the same stellar mass range used for our \gimic\ analysis from the 25 Mpc/h runs, resulting in 43 EAGLE galaxies for the REF-L0025N0376 run and 56 EAGLE galaxies for the RECAL-L0025N752 run, respectively.

In Fig.~\ref{fig:Mstar_GIMIC_EAGLE} we compare the present-day stellar mass$-$halo mass relations (left panel) and stellar mass$-$half-mass radius relations (right panel) of \gimic\ with the recent EAGLE simulations.  At stellar masses of less than approximately $10^{10}$ M$_\odot$, the simulations predict similar stellar mass$-$halo mass relations. Over this range of masses, the \gimic\ galaxies have slightly larger half-mass radii but there is significant overlap with the EAGLE galaxies.  At higher stellar masses, the \gimic\ simulations suffer from overcooling, resulting in higher stellar masses for a given halo mass and galaxies that are too compact (see also \citealt{mccarthy12b}).

\begin{figure}
\includegraphics[width=0.995\columnwidth]{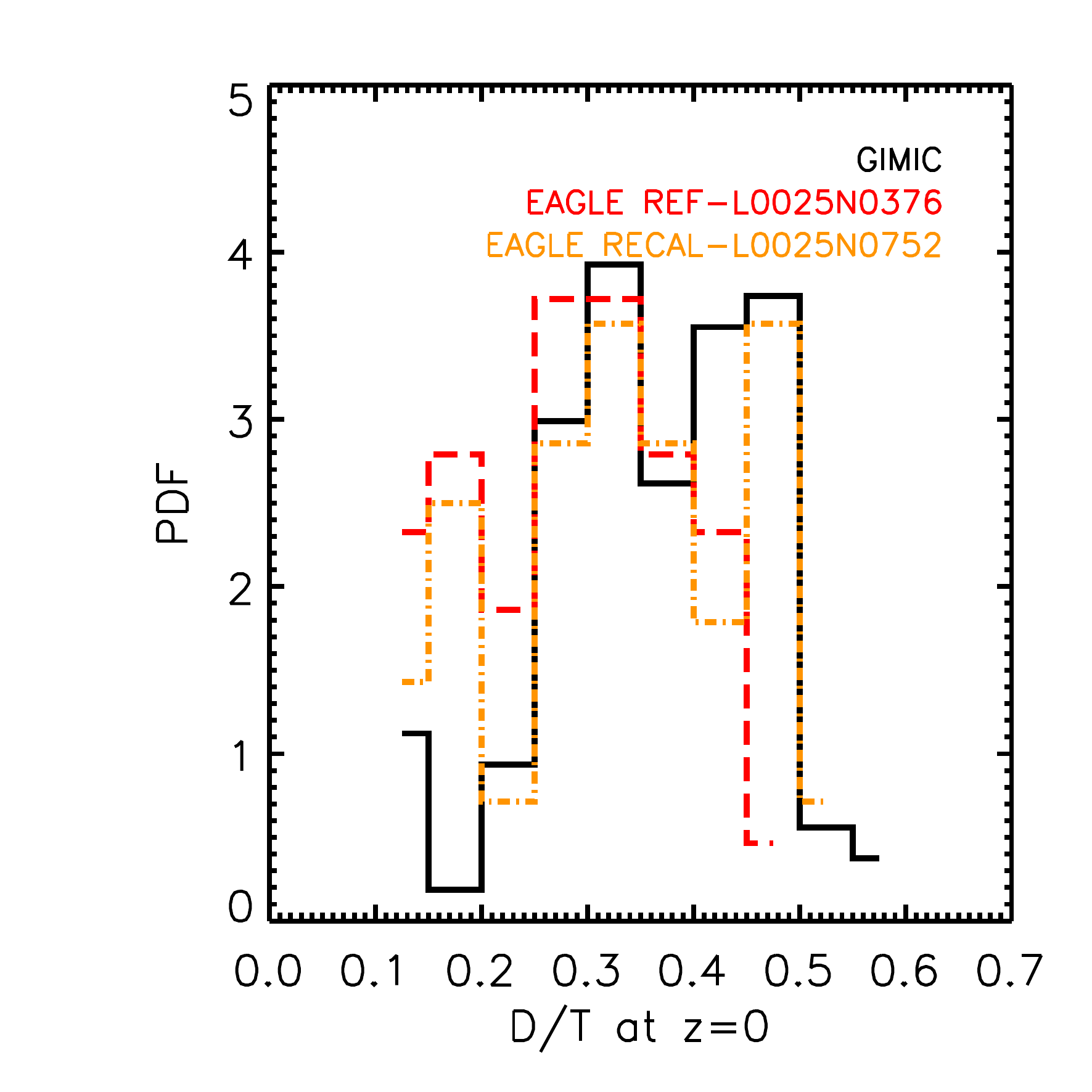} 
\caption{Comparison of the present-day D/T distribution of \gimic\ with the recent EAGLE simulations. The D/T distributions are very similar.}
\label{fig:eagle}
\end{figure}

We compare the D/T probability distribution functions of \gimic\ and the two EAGLE models in Fig.~\ref{fig:eagle}, where we have computed D/T in a consistent way for all of the simulations using the method described in the main text.  As can be seen, the three simulations have similar D/T distributions.

\section{Synthetic GAMA images}
\label{sec:appendix_gama}
To make a like-with-like comparison to the observational data, we produce synthetic GAMA r-band images of the simulated galaxies in both face-on and edge-on configurations.  Specifically, for each star particle in a simulated galaxy we estimate the (unabsorbed) r-band luminosity by treating it as a simple stellar population.  In particular, we use its initial stellar mass, metallicity, and age (and assume a Chabrier IMF) to interpolate a spectrum using the GALAXEV population synthesis package of \citet{bruzual03}.  We then apply the SDSS r-band transmission filter to the spectrum and then integrate it to obtain the r-band luminosity of the particle.  We place each simulated galaxy at a redshift of $z=0.1$ and we project the luminosities of all of the particles to produce two maps for each simulated galaxy (edge-on and face-on configurations), using a triangular-shaped-clouds interpolation scheme.  Each pixel has an angular size of 0.339 arcsec, corresponding to that adopted in the GAMA analysis, and each image has 101 pixels on a side, which spans a physical length of $\approx63$ kpc at the adopted redshift (assuming $H_0=70$ km/s/Mpc).  

At this stage we have flux images in units of ergs/s/cm$^2$.  We convert the flux images into counts images by multiplying by the SDSS exposure time of 53.9 seconds and the physical area of the telescope.  We further assume that the photons all have the same energy (corresponding to the effective wavelength of the r-band filter, 0.6231 microns) and that the gain (converting photons to electrons) is unity.  We then add a realistic sky component to the images (corresponding to 20.8 mag arcsec$^{-2}$).  Finally, we Poisson sample the galaxy+sky counts images and then subtract the sky (as done in GAMA data analysis).

The synthetic maps are convolved with the SDSS r-band point spread function (treated here as a Gaussian with a FWHM=1.1 arcsec) and then processed through the same S{\'e}rsic model fitting software applied to the real GAMA observational data (which uses the GALFIT software package of \citet{peng02} to do the actual fitting), providing us with estimates of the S{\'e}rsic indices of each of the simulated galaxies. The S{\'e}rsic model fitting yields estimates of the effective radius, the S{\'e}rsic index, the axial ratio, the position angle, the total magnitude, and the center coordinates.

\begin{figure*}
\centering
\includegraphics[width=0.65\columnwidth]{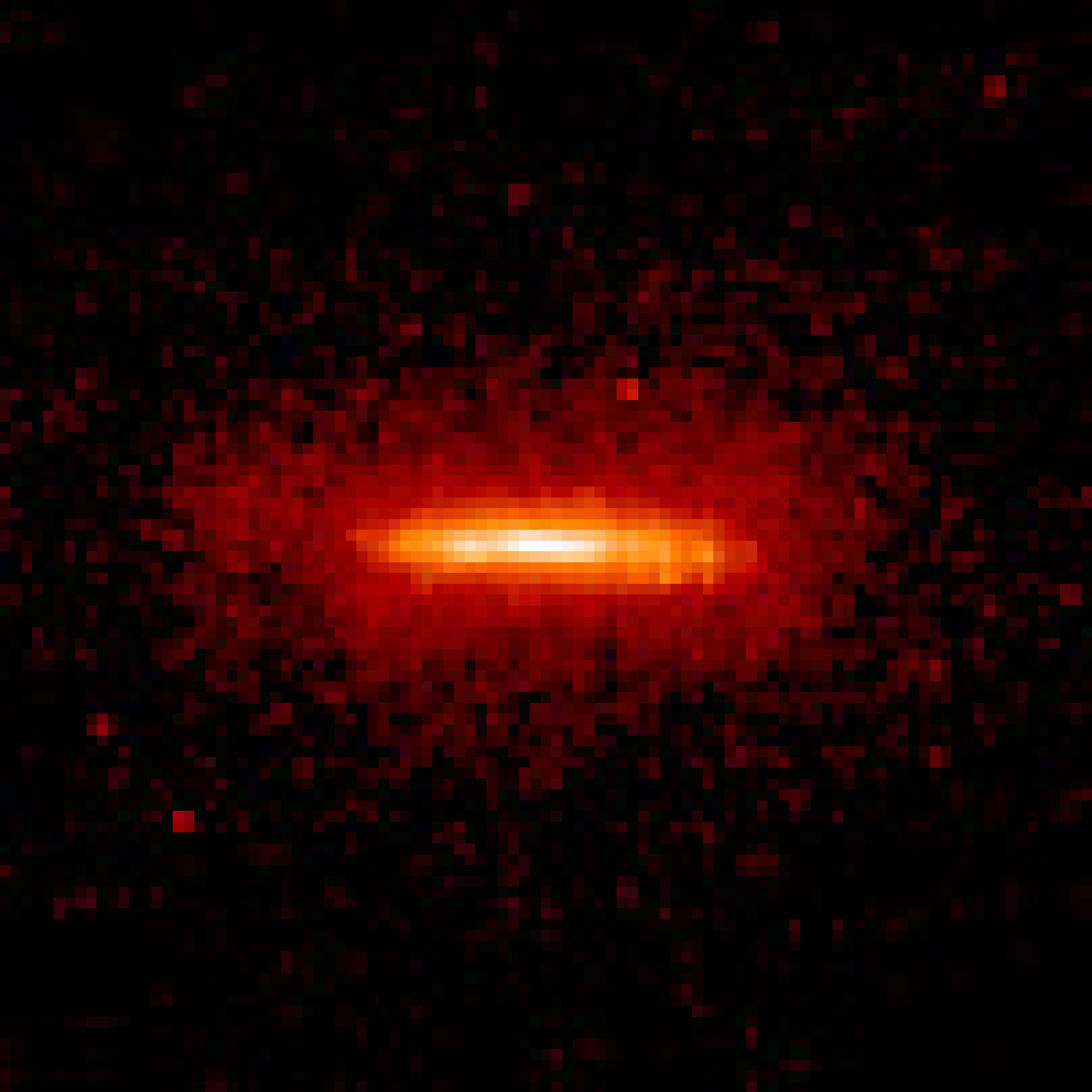}
\includegraphics[width=0.65\columnwidth]{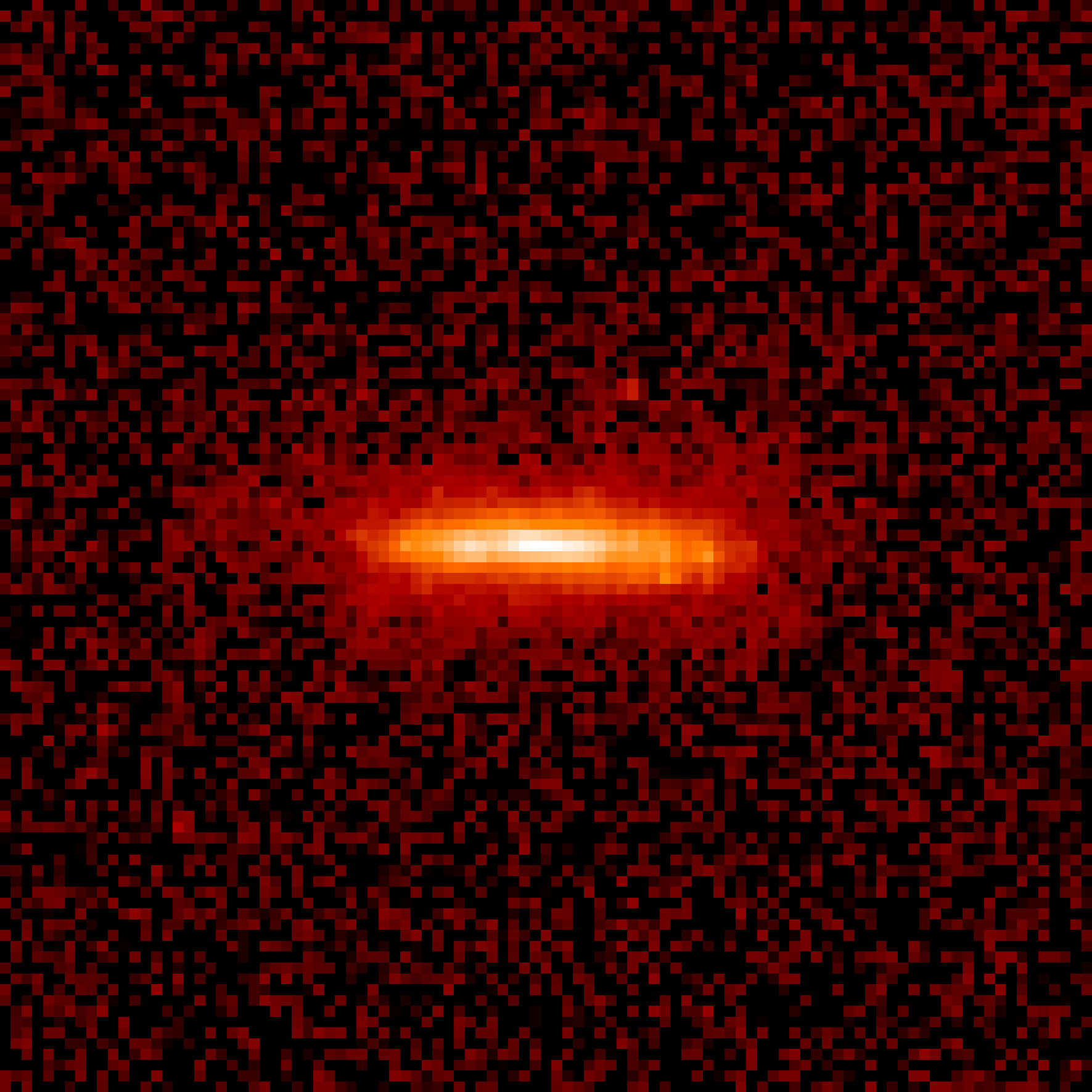}\\
\includegraphics[width=0.65\columnwidth]{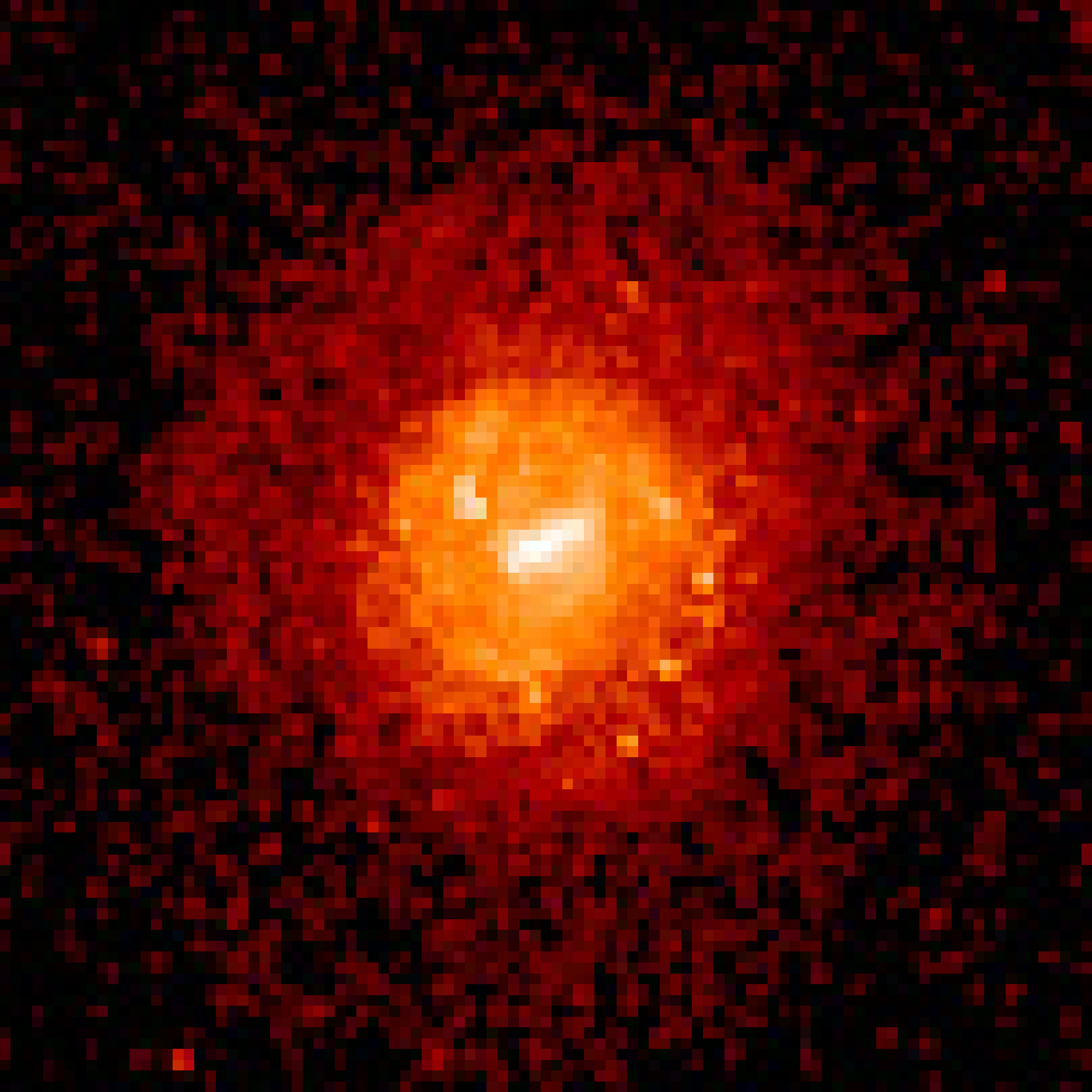}
\includegraphics[width=0.65\columnwidth]{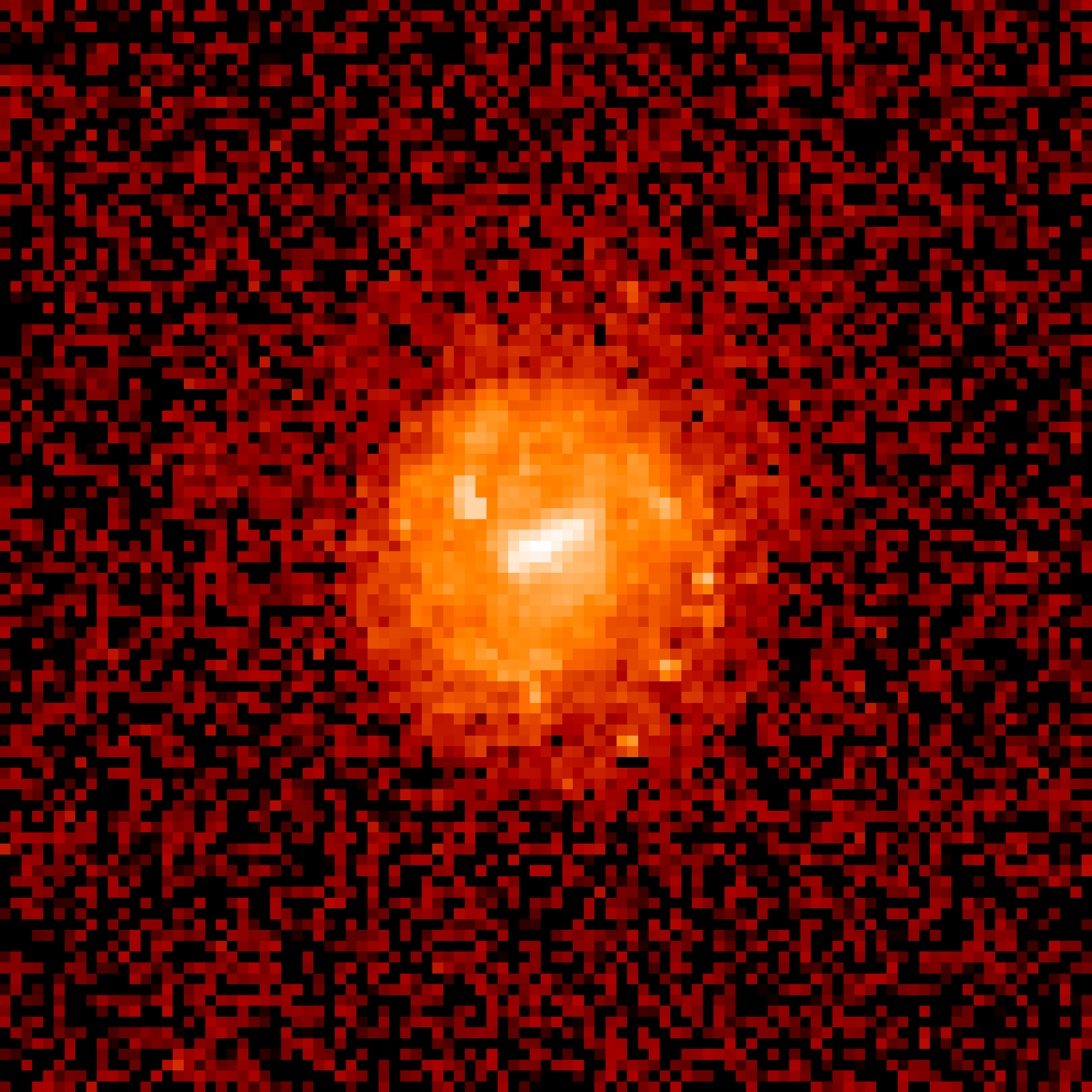}
\caption{Synthetic r-band counts images of a typical simulated galaxy, placed at $z=0.1$.  Each image has 101$^2$ pixels of length 0.339 arcsec, spanning a field of view of approximately 63 kpc.  The images are shown on a logarithmic scale, spanning a dynamic range of 4 orders of magnitude (i.e., white corresponds to the maximum and deep red corresponds to a counts level that is 4 orders of magnitude lower; black corresponds to an absence of counts).  The images in the top row show a show the galaxy in edge-on configuration, while the bottom row shows the galaxy in a face-on configuration.  The lefthand column shows the raw simulation images, while the righthand column shows the images after: i) a sky component was added; ii) the images were Poisson sampled; and iii) the sky was re-subtracted.  Note that the images have been convolved with the SDSS point spread function here.}
\label{fig:gal_images}
\end{figure*}

Here we provide an example from our synthetic GAMA imaging and surface brightness modelling pipeline.  Shown in Fig.~\ref{fig:gal_images} are virtual r-band counts images of a typical simulated galaxy, placed at $z=0.1$.  The images show the galaxy both in edge-on and face-on configurations and with and without Poisson sampling noise (see caption).  This particular galaxy has a strongly disky appearance with a hint of spiral structure when viewed in a face-on configuration.  The galaxy is sufficiently bright that the main surface brightness features are still clearly visible even when noise is added, although note that these images have not been convolved with the telescope point spread function.

\begin{figure*}
\centering
\includegraphics[width=0.995\textwidth]{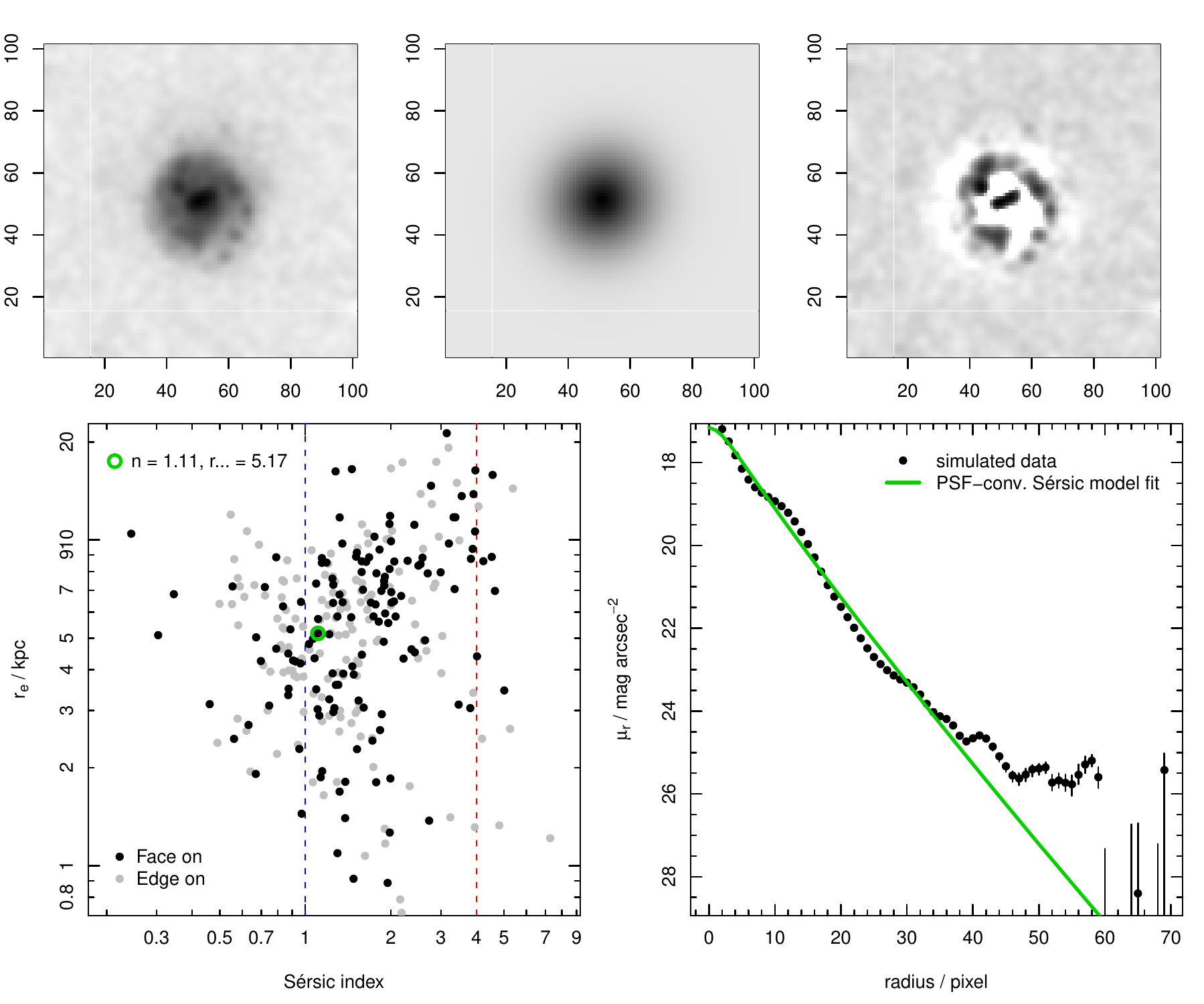}
\caption{Surface brightness modelling of the simulated galaxy shown in Fig.~\ref{fig:gal_images}.  The top row shows the noisy, PSF-convolved image of the galaxy in a face-on configuration (top left), the best-fit PSF-convolved 2D ellipsoidal S{\'e}rsic model (top middle), and the difference between the two previous images (top right).  The bottom row shows the best-fit S{\'e}rsic index and half-light radius of this galaxy (green circle) compared to the overall simulated population (bottom left) and the surface brightness profile of the galaxy along with the best-fit S{\'e}rsic model (bottom right).}
\label{fig:gal0_sbcomp}
\end{figure*}

In Fig.~\ref{fig:gal0_sbcomp} we show the best-fit 2D S{\'e}rsic model to the same simulated galaxy.  Overall, a single S{\'e}rsic profile with an index of approximately 1 (i.e., an exponential distribution) describes the surface brightness distribution of this simulated galaxy rather well, even if the model cannot reproduce the detailed spiral structure of the galaxy (nor would it do so for real galaxies with prominent spiral structure).

\end{appendix}

\end{document}